\documentclass[twocolumn]{aastex631}
\usepackage[dvipsnames]{xcolor}

\usepackage{amsmath}
\usepackage{array}
\usepackage{soul}
\begin{document}

\title{Modelling the Future of Gaia Neutron Star-Main Sequence Binaries: From Eccentric Orbits to Millisecond Pulsar-White Dwarfs}

\author[0000-0001-5867-5033]{Debatri Chattopadhyay}
\affiliation{Center for Interdisciplinary Exploration and Research in Astrophysics (CIERA) and Department of Physics \& Astronomy, Northwestern University, 1800 Sherman Ave, Evanston, IL 60201, USA}
\affiliation{NSF-Simons AI Institute for the Sky (SkAI), 172 E. Chestnut St., Chicago, IL 60611, USA}

\author[0000-0003-4474-6528]{Kyle A. Rocha}
\affiliation{Center for Interdisciplinary Exploration and Research in Astrophysics (CIERA) and Department of Physics \& Astronomy, Northwestern University, 1800 Sherman Ave, Evanston, IL 60201, USA}
\affiliation{Department of Astronomy \& Astrophysics, University of California, San Diego, 9500 Gilman Drive, La Jolla, CA 92093, USA}
\affiliation{NSF-Simons AI Institute for the Sky (SkAI), 172 E. Chestnut St., Chicago, IL 60611, USA}

\author[0000-0001-6692-6410]{Seth Gossage}
\affiliation{Center for Interdisciplinary Exploration and Research in Astrophysics (CIERA) and Department of Physics \& Astronomy, Northwestern University, 1800 Sherman Ave, Evanston, IL 60201, USA}
\affiliation{NSF-Simons AI Institute for the Sky (SkAI), 172 E. Chestnut St., Chicago, IL 60611, USA}

\author[0000-0003-1639-2929]{Vicky Kalogera}
\affiliation{Center for Interdisciplinary Exploration and Research in Astrophysics (CIERA) and Department of Physics \& Astronomy, Northwestern University, 1800 Sherman Ave, Evanston, IL 60201, USA}
\affiliation{NSF-Simons AI Institute for the Sky (SkAI), 172 E. Chestnut St., Chicago, IL 60611, USA}

\author[0000-0002-6871-1752]{Kareem El-Badry}
\affiliation{Department of Astronomy, California Institute of Technology, 1200 E. California Blvd., Pasadena, CA 91125, USA}

\author[0000-0002-9182-2047]{Alexander Tchekhovskoy}
\affiliation{Center for Interdisciplinary Exploration and Research in Astrophysics (CIERA) and Department of Physics \& Astronomy, Northwestern University, 1800 Sherman Ave, Evanston, IL 60201, USA}

\begin{abstract}

We model the future evolution of the 21 Gaia neutron star (NS) – main-sequence binaries (orbital period $P_{\mathrm{orb}}\sim200$–$1000\,\mathrm{days}$, eccentricity $e\gtrsim0.2$) using binary stellar evolution with \texttt{MESA}. We explore both models with eccentric mass transfer and traditional models that assume efficient circularization before the onset of mass transfer. All systems terminate their evolution as NS–white dwarf (WD) binaries, but different mass-transfer modes yield sharply different outcomes. Under eccentric mass transfer, binaries are driven to higher eccentricities, leading to most of the final orbits with $e\gtrsim 0.6$ and periods $P_{\rm orb} \sim 1000-4000$\,days. Periastron-triggered bursts are brief ($\lesssim10^6\,\mathrm{yr}$), transfer only a few $\times10^{-2}$\,M$_\odot$, and produce only mildly recycled pulsars (spin period $P_{\mathrm{spin}}\gtrsim50\,\mathrm{ms}$) with low-mass helium WDs. Artificially circularized mass transfer produces shorter final periods of $P_{\rm orb} \sim 200-2000$\,days and allows mass transfer to occur for $\sim10^7\,\mathrm{yr}$, so that the NSs accrete $\sim0.1$\,M$_\odot$. This yields fully recycled millisecond pulsars (MSPs) with $P_{\mathrm{spin}}\sim\mathrm{few}$–$30\,\mathrm{ms}$, including nine systems with carbon–oxygen WDs.  Allowing super-Eddington accretion up to $100\times$ the canonical limit makes even eccentric channels efficient recyclers that produce MSPs, though the torque coupling in this regime remains uncertain. Incorporating an adaptive, field-dependent magnetic-field decay time-scale for the very first time, we find that MSPs remain radio-alive over Gyr time-scales. All Gaia systems post stable mass transfer remains at large orbital period and fail to match the Galactic MSP-WD population where the bulk of the systems, mostly circular, appear at $P_{\rm orb}\lesssim100$\,days. Binaries with rather different mass ratio and initial orbital configuration (than the Gaia NS-MSs), which likely results to unstable mass transfer, are required to produce the bulk of the observed MSP-WD population. 

\end{abstract}

\keywords{Neutron stars — Millisecond pulsars — White dwarf stars — binary pulsars — Stellar evolution — Stellar accretion — Gaia}

\section{Introduction} \label{sec:intro}

The formation and evolution of neutron-star (NS) binaries provide a crucial window into high–energy astrophysics, stellar evolution, and the end stages of massive stars. These systems span a rich phenomenology—from X-ray binaries and double NS mergers to the recycled millisecond pulsars (MSPs) that power precision timing arrays. In particular, NSs paired with low-mass non-degenerate companions probe pathways that may lead to classical MSP - white dwarf (WD) systems, and inform models of mass-transfer, common-envelope (CE) evolution, and pulsar recycling.

Recent data from Gaia’s DR3 astrometry have revealed a population of NS - main-sequence (MS) binaries. From an initial sample of $\simeq1.7\times10^5$ stars—primarily ordinary solar-type main-sequence stars with periods of 1-3 years—precision photocentre orbits flagged 177 candidates with dark companions \citep{Shahaf2023}. Spectroscopic follow-up of a subset of these system has confirmed 21 high-probability NS-MS systems with dark companion masses $1.25-2$\,M$_\odot$ \citep{ElBadry2024}.These systems represent a new class of detached NS binaries, whose subsequent evolution has yet to be explored in detail.

In these 21 systems, the luminous primaries are typical low-mass stars with masses $M_1\simeq0.8-1.3$\,M$_\odot$, orbiting NS with masses $M_\mathrm{NS}\simeq1.3-1.9$\,M$_\odot$. The orbits have periods $P_\mathrm{orb}\simeq200-1000$\,days and unusually high eccentricities (e$\simeq0.1-0.8$) compared to WD-MS binaries, perhaps due to kicks during NS formation. These wide, eccentric systems—three of which harbour Li-enhanced, metal-poor halo stars—suggest potentially exotic formation channels \citep[e.g.][]{AndrewsPOSYDONv22024,Baibhav2025,Matsuno2025}. Their current orbital separations are large, suggesting that no prior mass transfer has occurred. In the standard binary-evolution pathway, the MS star eventually ascends the giant branch, initiates mass transfer onto the NS, loses its envelope, and leaves behind a WD remnant. Given the presently eccentric orbits, it remains uncertain whether these systems will circularize as the MS star evolves and fills its Roche lobe (RL), or instead undergo eccentric Roche-lobe overflow (RLOF), potentially forming eccentric NS–WD binaries. This evolutionary uncertainty—whether these systems circularize or maintain eccentric mass transfer—motivates the modelling presented in this work.

Radio surveys complement Gaia’s discoveries by revealing a handful ($\approx8$) of NS binaries with putative MS companions over the past few decades. Their inferred companion masses remain uncertain, spanning $\approx1$–$20$\,M$_\odot$, and the absence of clear accretion signatures suggests they are relatively young, non-recycled systems rather than products of prior mass transfer \citep[e.g.][]{Johnston1992ApJ,Kaspi1994ApJ,Champion2008Sci,Lyne2015MNRAS}. These systems likely represent an earlier, pre-mass-transfer (and hence pre-recycling) stage of binary evolution.  

As mass transfer begins or becomes sustained, systems can transition into the symbiotic X-ray binary (SyXB) phase, where the NS accretes from the wind of a $\sim1$\,M$_\odot$ red giant. Two benchmark systems in the period range relevant to our sample are IGR~J16194$-$2810, with $P_{\rm orb}\simeq193$\,days and $e\lesssim0.02$ \citep{Nagarajan2024,Hinkle2024}, and GX~1+4, with $P_{\rm orb}\simeq1160$\,days and $e\simeq0.1$ \citep{Hinkle2006}. These systems represent an intermediate evolutionary stage between detached NS–MS binaries and the fully recycled pulsars that eventually emerge as MSP–WD systems.

At the opposite end of the evolutionary sequence, radio surveys have long established the complementary population of recycled MSP–WD binaries. Roughly $\sim135$ known systems host helium WDs and $\sim43$ host carbon–oxygen (or oxygen–neon–magnesium) WDs, almost all in compact, nearly circular binaries ($e\lesssim10^{-3}$)\footnote{Only Galactic-field systems with measured median masses and orbital properties from the ATNF catalogue \citep{Manchester2005} are included, excluding globular-cluster pulsars; see \url{https://www3.mpifr-bonn.mpg.de/staff/pfreire/GCpsr.html}}. The contrast between Gaia’s wide, eccentric NS–MS binaries and the radio-discovered circular MSP–WDs raises the question of whether these two observed classes can be linked within a single evolutionary framework.

The evolutionary connection between these populations remains unclear. While reconstructing their past history is challenging, predicting the future evolution of the Gaia NS systems—despite our limited but steadily improving understanding of binary evolution—offers a tractable path forward. In this study, we aim to predict their future evolution (toward MSP–WD endpoints) using self-consistent stellar models in \texttt{MESA}. We extend standard binary evolution treatment by implementing a) the eccentric mass-transfer formalism of \citet{Sepinsky2007b, Sepinsky2009}, which allows both orbital separation $a$ and eccentricity $e$ to grow or decay depending on mass and angular-momentum losses, and b) by incorporating the pulsar spin evolution of \citet{Chattopadhyay2020} into \texttt{MESA}. This dual approach captures the full lifecycle of pulsar formation and evolution, from birth spins and kicks, through X-ray luminous accretion, to the emergence of a recycled MSP.

Our work combines Gaia and radio pulsar observations, self-consistent \texttt{MESA} binary evolution, and pulsar spin modelling to provide an integrated framework for NS–MS binaries. By linking eccentric mass-transfer physics to spin evolution, we generate direct predictions for pulsar timing observables—spin periods, orbital eccentricities, and binary demographics—that can be tested with current and future telescopes. Crucially, future surveys will expand discovery space: Gaia’s later data releases will deliver higher-precision orbits and fainter companions \citep{Brandt2024PASP,ElBadry2024NewA,An2025, Nagarajan2025PASP..137d4202N}, while new radio searches with MeerKAT and the SKA will be uniquely sensitive to long-period, eccentric pulsars that current algorithms often miss \citep{2015aska.confE..40K, 2015aska.confE..39T, Meer2023A&A,Meer2023MNRAS}. Although it remains uncertain whether Gaia’s eccentric NS–MS binaries evolve into the circular NS-WD systems known from radio pulsar surveys, our framework provides the first self-consistent test of this connection. 

We organize this paper as follows. In Section~\ref{sec:methods}, we describe our methodology and model setup. In Section~\ref{sec:results}, we present our results, focusing on detailed evolutionary tracks in Section~\ref{sec:futureMESA}, the final orbital properties in Section~\ref{sec:porb_e}, the final WDs (mass and type) from the evolving MS stars in Section~\ref{sec:WD}, and the
the associated Hertzsprung–Russell diagram in Section~\ref{sec:hr}. In Section~\ref{sec:wd_pulsars} we discuss in detail the spin and surface magnetic-field evolution of the NSs as pulsars. Finally, in Section~\ref{sec:summary} we summarize our conclusions and outline prospects for future work.

\begin{table*}
\centering
\begin{tabular}{lrrrrrrrrrrrr}
\toprule
     names & ID &   $P_{\rm orb}$ &     $a$ &      $e$ &   $M_\mathrm{NS}$ & $M_\mathrm{1}$ & $R_\mathrm{1}$ &  $T_\mathrm{eff}$ & $Z$ & $Z_{\rm grid}$ &  $t_{\rm 1,o}$  & $t_{\rm 1,std}$ \\
     &&(days)&(AU)&&(M$_\odot$)&(M$_\odot$)&(R$_\odot$)&(kK)&&(/$Z_\odot$)&Gyr&Gyr\\
\hline
J0553-1349 & 0 & 189.1 & 0.851 & 0.3879 & 1.33 & 0.98 & 0.988 & 5.59 & 0.018442 & 1 & 6.79 & 2.26 \\
J2057-4742 & 1 & 230.15 & 0.978 & 0.3095 & 1.31 & 1.048 & 0.941 & 5.834 & 0.0188446 & 1 & 3.19 & 1.45 \\
J1553-6846 & 2 & 310.17 & 1.194 & 0.5314 & 1.323 & 1.04 & 0.971 & 5.88 & 0.0188446 & 1 & 4.12 & 1.14 \\
J0742-4749 & 3 & 497.6 & 1.593 & 0.168 & 1.28 & 0.9 & 1.259 & 5.68 & 0.0106568 & 1 &  11.32 & 1.43 \\
J0003-5604 & 4 & 561.83 & 1.717 & 0.795 & 1.34 & 0.802 & 0.752 & 4.885 & 0.0116452 & 1 & 2.07 & 1.08 \\
J1733+5808 & 5 & 570.94 & 1.835 & 0.3093 & 1.362 & 1.16 & 1.081 & 6.01 & 0.0205385 & 1 & 3.48 & 1.93 \\
J0217-7541 & 6 & 636.1 & 1.936 & 0.3228 & 1.396 & 0.996 & 1.238 & 5.58 & 0.0192555 & 1  & 11.06 & 1.02 \\
J0639-3655 & 7 & 654.6 & 2.132 & 0.721 & 1.7 & 1.32 & 1.43 & 6.48 & 0.0154994 & 1 &  1.75 & 1.28 \\
J1048+6547 & 8 & 827 & 2.344 & 0.357 & 1.52 & 0.99 & 1.15 & 5.96 & 0.0104227 & 1 &  6.78 & 2.34 \\
J2145+2837 & 9 & 889.5 & 2.405 & 0.584 & 1.396 & 0.95 & 0.846 & 5.5 & 0.0132941 & 1 &  4.09 & 2.25 \\
J2244-2236 & 10 & 938.3 & 2.527 & 0.5666 & 1.443 & 1.002 & 0.881 & 5.892 & 0.0161905 & 1 & 4.10 & 2.19 \\
J0230+5950 & 11 & 1029 & 2.713 & 0.753 & 1.401 & 1.114 & 1.001 & 5.87 & 0.0196746 & 1 & 0.23 & 0.46 \\
J0634+6256 & 12 & 1046 & 2.79 & 0.564 & 1.48 & 1.18 & 1.68 & 6.05 & 0.0132941 & 1 &  2.41 & 1.13 \\
J0152-2049 & 13 & 536.14 & 1.647 & 0.6615 & 1.291 & 0.782 & 1.169 & 6.47 & 0.000780146 & 0.01 & 12.86 & 0.45 \\
J1739+4502 & 14 & 657.4 & 1.914 & 0.6777 & 1.38 & 0.781 & 1.424 & 6.289 & 0.000225413 & 0.01 &  12.77 & 3.37 \\
J1432-1021 & 15 & 730.9 & 2.208 & 0.1203 & 1.898 & 0.79 & 1.491 & 6.049 & 0.000762432 & 0.01 & 13.30 & 0.20 \\
J1449+6919 & 16 & 632.65 & 1.868 & 0.2668 & 1.258 & 0.91 & 1.06 & 6.03 & 0.00329997 & 0.2 &  7.67 & 2.75 \\
J2102+3703 & 17 & 481.04 & 1.632 & 0.448 & 1.473 & 1.033 & 0.947 & 6.27 & 0.00622608 & 0.45  & 1.14 & 1.51 \\
J0036-0932 & 18 & 719.8 & 2.075 & 0.3993 & 1.362 & 0.94 & 1.32 & 5.84 & 0.00556153 & 0.45 & 10.51 & 1.59 \\
J1150-2203 & 19 & 631.81 & 1.973 & 0.552 & 1.39 & 1.18 & 1.495 & 5.834 & 0.0346791 & 2 & 2.52 & 0.50 \\
J0824+5254 & 20 & 1026.7 & 2.776 & 0.686 & 1.604 & 1.102 & 0.997 & 5.89 & 0.025407 & 2 &  2.28 & 4.81 \\
\hline
\end{tabular}
\caption{Initial orbital and stellar parameters for the 21 \textit{Gaia} systems used to initialize our \texttt{MESA} runs from \protect{\cite{ElBadry2024}}. From left to right, columns give: (1) system identifier; (2) binary index; (3) orbital period $P_{\rm orb}$ (days); (4) semi‐major axis $a$ (AU); (5) eccentricity $e$; (6) neutron‐star mass $M_{\rm NS}$ (M$_\odot$); (7) companion mass $M_{\rm 1}$ (M$_\odot$); (8) companion radius $R_{\rm 1}$ ($R_\odot$); (9) companion effective temperature $T_{\rm eff}$ (kK); (10) spectroscopic metallicity $Z$; (11) chosen metallicity grid $Z_{\rm grid}$ (nearest available, in units of stellar metallicity $Z_\odot$); (12) companion age $t_{\rm 1,o}$ (Gyr) estimated from SED fitting; and (13) uncertainty in isochrone age $t_{\rm 1,std}$ (Gyr). Measurement errors are \(1\sigma\), typical precisions (median of the quoted errors) are: $P_{\rm orb} \pm$\,0.6\,d (0.09\%), $M_{\rm NS} \pm$\,0.04\,M$_\odot$ (2.9\%), $M_{\rm 1} \pm$\,0.04\,M$_\odot$ (4\%) and $e \pm$\,0.002 (0.36\%).}
\label{Tab:GaiaNSBinaries}
\end{table*}

\section{Methods} \label{sec:methods}
 
We investigate the future evolution of Gaia-identified NS–MS binaries using self-consistent stellar models in \texttt{MESA} \citep{Paxton2011}. Since the Gaia NS binaries show high eccentricity, our models incorporate the eccentric mass-transfer prescriptions of \citet{Sepinsky2007b, Sepinsky2009} together with the pulsar spin evolution of \citet{Chattopadhyay2020}, allowing us to track magnetic-dipole spin-down, magnetic-field decay, and accretion-driven spin-up during mass-transfer phases. The detailed methodology is presented in this section. We focus on predicting the future evolution of the 21 Gaia NS–MS binaries and exploring their potential connections to the observed Galactic radio pulsar–WD population.

\subsection{MESA}\label{sec:mesa}

We use version r11701 of the \texttt{MESA} one-dimensional stellar evolution code \citep{Paxton2011,Paxton2013,Paxton2015,Paxton2018,Paxton2019}, which computes detailed evolutionary tracks with adaptive mesh refinement and timestep control, modern equation of state and opacity tables, extensive nuclear reaction networks, and parametrized treatments of convection, rotation, and mass loss. We apply this framework to model the evolution of the Gaia NS–MS binaries (as described in \citealt{Fragos2023,AndrewsPOSYDONv22024}).

Our \texttt{MESA} models include the magnetic braking prescription of \citet{Garraffo2018}, modeling the angular-momentum loss as
\[
\begin{aligned}
\dot{J} &= \dot{J}_{\rm dip}\,Q_J(n),\\
\dot{J}_{\rm dip} &= -c\,\Omega^3\,\tau_c,\\
Q_J(n) &= 4.05\exp\bigl(-1.4\,n\bigr).
\end{aligned}
\]

where the magnetic complexity number 
\begin{equation}
n = a\,{\rm Ro}^{-1} + b\,{\rm Ro} + 1
\end{equation}
is parameterized via the Rossby number ${\rm Ro}=P/\tau_c$. In the above equations, $\Omega$ represents the (equatorial) angular rotation rate of the star; $\tau_c$ is its convective turnover time (measured one pressure scale height from the base of the convection zone); $P$ is the rotation period of the star; and $n$ parameterizes the magnetic field complexity of the star, with a value of $1$ corresponding to a dipole configuration, $2$ a quadrupole, and so on. In this formalism, dipolar configurations lead to more efficient magnetic braking and higher order magnetic field configurations yield weaker efficiency. In representing $n$, $a$ and $b$ are free parameters, calibrated as in \cite{Gossage2023}. This exponential modulation of angular momentum loss produces a rapid transition between `saturated' and `unsaturated' braking regimes, reproducing both the bifurcated rotation-period distributions observed in open clusters and the strong braking needed to form ultra-compact NS low-mass X-ray binaries (LMXBs). Thus, this braking formalism is capable of consistently reproducing the spin down of single stars and the rotational/orbital evolution of LMXBs, the suspected progenitor systems of millisecond pulsars. We implement this in \texttt{MESA} using the \citet{Gossage2023} update to the \citet{Garraffo2018} formulation.

Instead of using the standard circularization RLOF prescription often adopted in binary stellar evolution calculations \citep{Eggleton1983}, we self-consistently model the orbital evolution through eccentric mass transfer using the formalism by \cite{Sepinsky2007b, Sepinsky2009} as implemented in \texttt{MESA} by \cite{Rocha2025}.
In our fiducial model, we replace the standard circularized RLOF prescription \citep{Eggleton1983}. These analytic secular-evolution equations assume periastron-only ($\delta$-function) mass transfer and provide orbit-averaged expressions for the changes in semi-major axis, $\langle\dot{a}\rangle$, and eccentricity, $\langle\dot{e}\rangle$, in terms of the instantaneous periastron transfer rate $\dot{M}_0$. 
These expressions are implemented in the \texttt{MESA} binary module by replacing the circular RL torque with the eccentric mass transfer contributions paramterized through the total orbital angular‐momentum budget:
\begin{equation}
\dot J_{\rm orb} \;=\;\dot J_{\rm gr} + \dot J_{\rm ml} + \dot J_{\rm mb} + \dot J_{\rm ls}\,,
\end{equation}
where within $\dot J_{\rm ml}$ we substitute the circular RL term with the eccentric mass-transfer term from \cite{Sepinsky2009}, and compute the periastron‐peak rate $\dot{M}_0$ using the orbit‐averaged prescription of \cite{Ritter1988} with the \cite{Eggleton1983} Roche geometry (Eq.~7).  

We treat non-conservative mass-transfer with an accretion efficiency $\beta$, so that $\dot M_\mathrm{NS} = -\beta\,\dot M_1$, capping accretion at the Eddington rate and ejecting any excess with the NS accretor’s specific orbital angular momentum \citep{Sepinsky2007b,Sepinsky2009}. We also consider a fully conservative limit $\beta=1$, i.e., all transferred mass is accreted—a deliberate  upper bound on spin-up and orbital response. We refer to these two cases as the `Eddington-limited' and `super-Eddington conservative' models, respectively. Finally, we consider a super-Eddington, non-conservative case in which accretion may exceed the Eddington rate by up to a factor of 100. 

Finally, we evolve the orbit by adding the eccentric mass-transfer contributions to $\dot a$ and $\dot e$ together with gravitational-radiation \citep{Peters1964}, stellar-wind, magnetic-braking \citep{Garraffo2018}, and tidal-coupling terms to evolve the stellar structure and orbital parameters self-consistently \citep{Paxton2018,Paxton2019,Paxton2011}.

\begin{table*}[ht]
\centering
\begin{tabular}{@{}lcc@{}}
\toprule
\textbf{Model} & \textbf{Orbit Geometry}     & \textbf{MT Regime}         \\
\hline
\texttt{e-Ed}      & Eccentric                & Eddington-limited      \\
\texttt{e-SpEd100}
   & Eccentric                & Super-Eddington ($\times 100$ capped)       \\
\texttt{e-SpEdcn}
   & Eccentric                & Super-Eddington (conservative)       \\
\texttt{cir-Ed}    & Circularized at RLOF     & Eddington-limited      \\
\hline
\end{tabular}
\caption{Summary of \texttt{MESA} binary model variants. `MT' signifies mass transfer.  We also modelled super-Eddington runs at $2\times$, $5\times$, and $500\times$; the former two were similar to \texttt{e-Ed}, and the latter was nearly indistinguishable from \texttt{e-SpEdcn}, so they are not shown separately. We also tested circularization at the observed orbit, but its results were indistinguishable from those of the \texttt{cir-Ed} case, so they are not shown separately.}

\label{Tab:modelvariants}
\end{table*}

In total, for each of the 21 Gaia systems we construct four \texttt{MESA} models (Table~\ref{Tab:modelvariants}): (i) eccentric mass-transfer with Eddington-limited accretion (\texttt{e-Ed}), (ii) eccentric mass-transfer with fully conservative super-Eddington accretion (\texttt{e-SpEdcn}), (iii) eccentric mass-transfer super-Eddington accretion limited at 100$\times$ over the Eddington limit (\texttt{e-SpEd100}) and (iv) a circularized analogue at the onset of RLOF with Eddington-limited accretion (\texttt{cir-Ed}). Although this setup assumes instantaneous, artificial circularization at the onset of RLOF, it remains the standard baseline adopted in most classical binary-evolution studies (e.g. \citealt{Hurley2002, Izzard2004, Tauris2012}). We include it here primarily as a control model to benchmark against the more physically consistent eccentric prescriptions, noting that several works have demonstrated that such forced circularization can yield systematically biased post-RLOF outcomes (e.g. \citealt{Sepinsky2007b,Lajoie2011,Davis2013,Hamers2019}). We also tested an alternative circularization at the current epoch, but its results are nearly indistinguishable from \texttt{cir-Ed} and are not discussed further. For the circularized case, we match the periastron separation, $r_p = a(1-e)$, setting $a_{\rm circ}=r_p$, and computing the corresponding circular period from Kepler’s law,
\begin{equation}
    P_{\rm circ} = 2\pi\sqrt{\frac{a_{\rm circ}^3}{G(M_1+M_2)}}
\end{equation}
with $e=0$.  These $(a_{\rm circ},P_{\rm circ},e=0)$ serve as the initial conditions in the \texttt{cir-Ed} runs. The alternative circularization yields essentially identical results because for $\simeq 1$\,M$_\odot$ stars, line-driven winds are intrinsically weak for such small masses (and even weaker at low metallicity). Other processes that might alter the orbit before RLOF, such as gravitational radiation and magnetic braking, are negligible for our wide binaries ($P_\mathrm{orb} \gtrsim 150$\,days). The resulting \texttt{MESA} model regimes are summarized in Table~\ref{Tab:modelvariants}.

The initial conditions for our simulations—component masses, orbital periods, and, for eccentric models, eccentricities—are drawn to match the observed properties of the 21 Gaia binaries as shown in Table~\ref{Tab:GaiaNSBinaries}, (with metallicites chosen to match the resolution limits of ZAMS models available in \citealt{AndrewsPOSYDONv22024}). Companion ages were adopted from the SED (spectral energy distribution) fits reported in Section\,4 of \citet{ElBadry2024}\footnote{study derived stellar masses, radii, and temperatures; the companion ages listed in Table~\ref{Tab:GaiaNSBinaries} are by-products of their SED fits and were not separately published}. These ages are based on single-star models, an appropriate simplification since none of the systems shows evidence of prior RLOF. The inferred companion ages ($t_{1,o}$; Table~\ref{Tab:GaiaNSBinaries}, column 12) have large $1\sigma$ uncertainties ($t_{1,\mathrm{std}}$; column 13), often comparable to their mean values—reflecting the well-known difficulty of age-dating MS stars. In our simulations, we initiate the tidal evolution at the observed companion age $t_{1,o}$, but note that these uncertainties have no impact to our conclusions.

Given the low stellar masses and wide orbital separations, tidal interactions remain weak until the companion expands off the MS and approaches RLOF. Consequently, the precise age of the MS star does not affect the subsequent orbital or mass-transfer evolution, and our results are robust against these uncertainties.

Beyond tides, the age offset affects only our predictions for pulsar spin periods—and even then, only weakly—because pulsar spin-down is primarily governed by prior mass accretion and angular momentum gain, rather than absolute system age. Moreover, future mass accretion depends on orbital evolution, which is itself driven by present-day masses, radii, and orbital parameters. In conclusion, uncertainties in companion age have negligible influence on our results.

\subsection{Pulsar modelling}\label{sec:pulsar}
We follow the prescriptions as developed in \cite{Chattopadhyay2020, Chattopadhyay2021} for modelling the evolution of NS spin and surface magnetic field. In an `isolated' scenario, i.e., one with no mass-transfer, the NS is considered to be born as a magnetised, rapidly rotating object that spins down by losing angular momentum via magnetic dipole radiation (i.e., as a rotationally powered pulsar). Then, the rate of change of the spin angular frequency is calculated as 
\begin{equation}\label{eqn:isolatedSpin}
    \dot{\Omega}=-\frac{8\pi B^2 R^6 \sin^2\alpha \Omega^3}{3\mu_0 c^3 I},
\end{equation}
where $\Omega=2\pi/P_\mathrm{spin}$ ($P_\mathrm{spin}$ being the spin period) is the angular frequency, $B$ is the surface magnetic field strength, $R$ is the radius, $\alpha$ is the angle between the spin and the magnetic axis, and $I$ is the moment of inertia \citep[Equation\,12 from][]{Lattimer2005} of the NS.
The constants $c$ and $\mu_0$ are the speed of light in vacuum and free space permeability, respectively. Equation~\ref{eqn:isolatedSpin} follows $\dot{\Omega}\propto\Omega^b$, with $b=3$, the exponent $b$ is commonly called the pulsar braking index. Most stable, recycled pulsars are observed to have a braking index of $\simeq$3, in support of dipolar magnetic field \citep{Lorimer2004book,Kiziltan2010}, although young pulsars often show $b<3$ due to additional torques (e.g. winds, evolving fields), and in rare transient episodes (e.g. post-glitch relaxation) measured values can momentarily reach $b\gg3$ (even $\sim$90), without reflecting the secular spin-down law \citep{Espinoza2011, Archibald2016}. We adopt a simple exponential surface magnetic field decay, as expected from ohmic dissipation (with Hall/ambipolar contributions) and supported by population studies indicating secular field decay in radio pulsars \citep{Goldreich1992,Igoshev2014}. The surface magnetic field strength in the isolated case is modelled to be decaying exponentially over time as 
\begin{equation}\label{eqn:isolatedMagneticField}
B=(B_0-B_\mathrm{min})\times \exp{(-t/\tau)} + B_\mathrm{min},
\end{equation}
where $B_0$ is the initial surface magnetic-field strength, $B_\mathrm{min}=10^8\,$G is the floor set by accretion-induced magnetic burial and crustal (diamagnetic) screening—consistent with the residual dipole strengths observed in recycled millisecond pulsars, for which further large-scale decay appears to stall \citep{Bhattacharya1991,Cumming2001,Payne2007}, $t$ is time, and $\tau$ is the `magnetic field decay time scale'. Equations~\ref{eqn:isolatedSpin} and \ref{eqn:isolatedMagneticField} are solved to calculate $\Omega$ as a function of time \citep[see equation 6 of][]{Chattopadhyay2020}. In our spin modelling, we subtract the companion's age (see Section~\ref{sec:mesa}) from the pulsar’s characteristic age to form a `shifted' age (that is, accounting for the time since the NS's birth to its current spin-down state).  We find that using either the raw or shifted age changes the predicted spin period by $\lesssim 5\%$, an uncertainty small compared to our other uncertainties including the NS birth spin and magnetic field, as well as $\tau$.

Traditional binary-population studies typically assign a single, field-independent decay constant $\tau$ to every neutron star in the simulation (e.g. \citealt{Kiel2008, Kiel2010, Chattopadhyay2020, Chattopadhyay2021, Sgalletta2023}). Observations, however, reveal a more nuanced picture: young pulsars with surface fields in the range $B \sim 10^{11\text{–}12}$\,G fade on Myr–Gyr time-scales \citep{Igoshev2015}, whereas recycled MSPs (spun up by prior accretion in a low-mass X-ray binary) with $B \sim 10^{8\text{–}9}$\,G retain nearly constant fields over a Hubble time \citep{Bell1995,Hansen1998,Johnston1999}. Radio pulsars with $B \lesssim 10^8$\,G are not observed, likely because they fall below the pair-production death line—i.e., they can no longer generate sufficient electron–positron pairs to sustain coherent radio emission \citep{Zhang2000, Hibschman2001}. 
Magneto-thermal calculations attribute this contrast to two crustal processes. Hall drift primarily redistributes magnetic flux and acts on a time-scale that is inversely proportional to $B$, Ohmic diffusion dissipates the currents, and its time-scale lengthens as the crust cools and the field weakens \citep{Sengupta1997, TauD2009A&A...496..207P,TauD2013MNRAS.434..123V, TauD2021Univ....7..351I}. Taken together, these processes motivate $\tau\propto B^{-1}$ to first order. To reflect this physical picture, we replace the single constant $\tau$ with a field-dependent value once the star has been recycled. Immediately after accretion ceases (i.e., at the end of RLOF), we set, post mass-accretion recycled pulsars, \begin{equation}\label{equ:tau}
    \tau_\mathrm{r}=\tau_0\,
    \left( \frac{B_{\mathrm r}}{B_{0}} \right)^{-\eta},
\end{equation}
where $\tau_0$ and $B_0$ are the initial magnetic field decay time-scale and birth surface magnetic field strength of the pulsar, while $B_\mathrm{r}$ is the surface field strength at the end of mass-transfer after recycling, and $\eta\simeq$0.8-1 \citep[guided by][we set $\eta=1$ in our simulations]{Sengupta1997,Geppert2002,TauD2013MNRAS.434..123V}. This scaling lengthens the decay time by two-three orders of magnitude for a typical MSP, reproducing both the rapid decay of high-$B$ young pulsars and the near-freezing of fields in recycled systems without introducing additional free parameters. This behaviour is consistent with magneto-thermal theory---Hall drift ($t_{\rm Hall}\!\propto\!1/B$) and cooling-controlled Ohmic diffusion both slow sharply as $B$ decreases---so lowering $B$ from $10^{11\text{--}12}$\,G (young pulsars) to $10^{8\text{--}9}$\,G (MSPs) naturally lengthens the effective decay time by $\sim 10^{2\text{--}3}$, matching rapid early decay and near-freezing in recycled systems \citep{Gourgouliatos2014, Pons2009,  Igoshev2015, Bell1995, Hansen1998, Johnston1999,Cumming2001}.

During mass-transfer due to RLOF the pulsar is expected to accrete matter that buries its surface magnetic field as well as spins it up \citep{JahanMiri,Zhang2006,Kiel2008}. We calculate the rate of change of angular velocity due to accretion \citep{Ghosh1979,Ghosh1979ApJ} as 
\begin{equation}\label{eqn:accSpin}
\dot{\Omega}_{\mathrm{acc}}= \frac{\epsilon\,V_{\mathrm{diff}}\,R_{\mathrm{M}}^{2}\,\dot{M}_{\mathrm{NS}}}{I}
\end{equation}
and the evolution of the surface magnetic field as a result of the accreted matter \citep{Shibazaki1989, Konar1997}
\begin{equation}\label{eqn:accMagneticField}
B_\mathrm{acc}=(B_0-B_\mathrm{min})\times \exp{(-\Delta M_\mathrm{NS}/\kappa)} + B_\mathrm{min},
\end{equation}
where $\epsilon$ is the efficiency factor, $V_\mathrm{diff}$ is the difference between the Keplerian angular velocity evaluated at the magnetic radius $R_\mathrm{M}(=R_\mathrm{Alfven}/2,\text{i.e. half of the Alfven radius})$ and the angular velocity at which the NS magnetosphere corotates, $\Delta M_\mathrm{NS}$ is the mass accreted by the NS and $\kappa$ is the `magnetic field decay mass-scale' \citep[$\kappa \sim 0.01–0.1$\,M$_\odot$, from][]{Shibazaki1989, Konar1997}. 
\begin{figure*}[!htbp]
  \centering
  \includegraphics[width=0.5\textwidth]{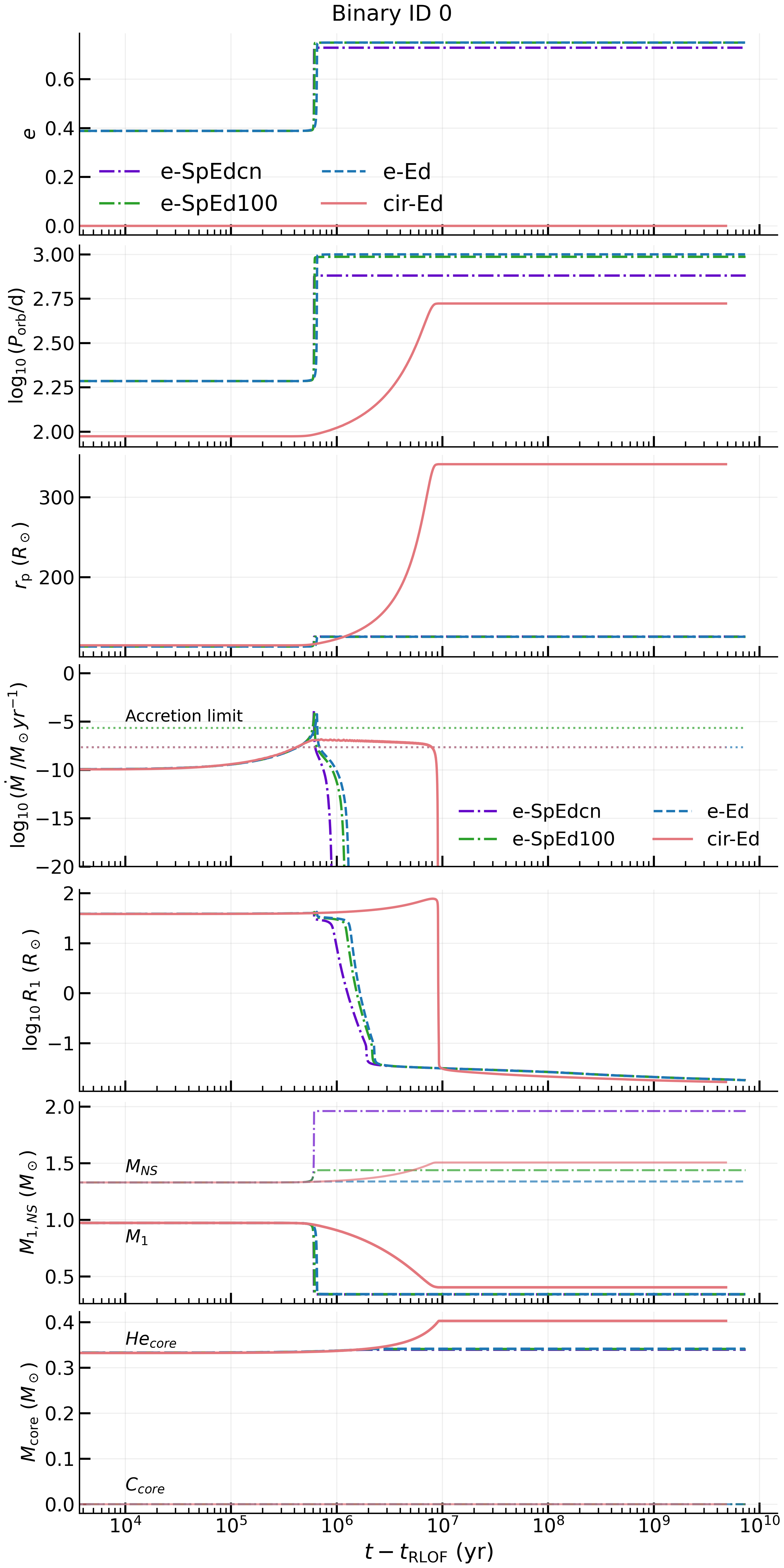}%
  \includegraphics[width=0.5\textwidth]{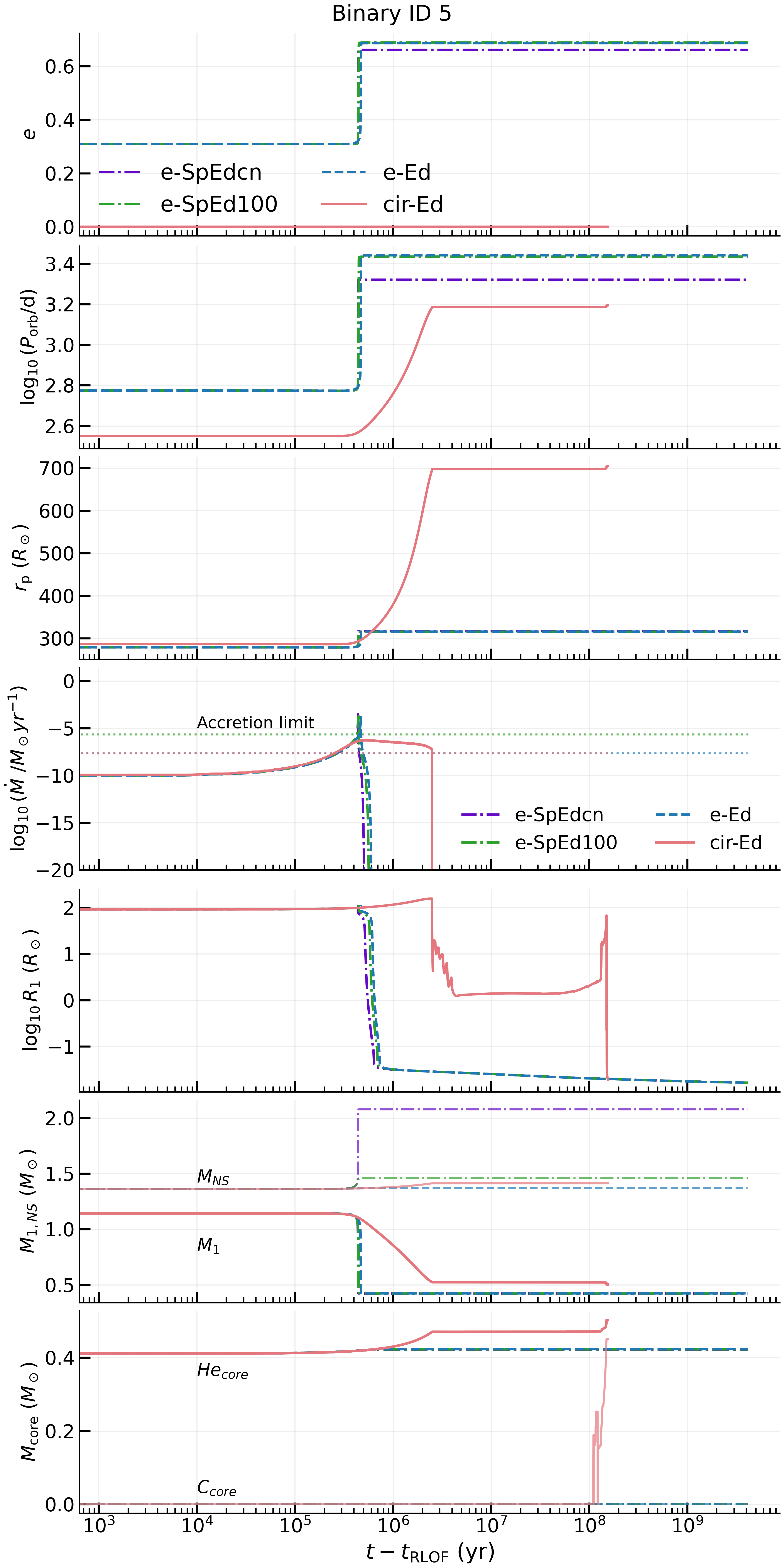}%
\caption{Time evolution of key binary and stellar properties for binary ID 0 (left column) and binary ID 5 (right column). We compare four mass-transfer scenarios: Eddington-limited circular (\texttt{\detokenize{cir–Ed}}, pink), Eddington-limited eccentric (\texttt{\detokenize{e–Ed}}, blue), conservative super-Eddington eccentric (\texttt{\detokenize{e–SpEdcn}}, purple), and eccentric with $100\times$ Eddington rate (\texttt{\detokenize{e–SpEd100}}, green).
Panels from top to bottom show: (a) orbital eccentricity $e$; (b) $\log_{10}(P_{\rm orb}/{\rm d})$; (c) orbital separation at periastron $r_\mathrm{p}$; (d) companion radius $\log_{10}(R_1/R_\odot)$; (e) masses of the companion $M_1$ (solid) and neutron star $M_2$ (dash–dotted); (f) mass-transfer rate $\log_{10}(\dot M/M_\odot,{\rm yr}^{-1})$; and (g) helium (solid) and carbon (dashed) core masses.
Time is measured relative to Roche-lobe overflow, $t - t_{\rm RLOF}$. Eccentric models produce short-lived, strongly peaked mass-transfer episodes that consistently form helium WDs, while extended transfer in \texttt{\detokenize{cir–Ed}} may occasionally yield carbon–oxygen WDs (e.g., binary ID 5), for our Gaia NS binaries. These binary-evolution outcomes directly feed into our pulsar modelling (Section~\ref{sec:pulsar}), where the mass-transfer histories determine the subsequent spin-up and magnetic-field burial of the NS.}
  \label{fig:timeSeries}
\end{figure*}
We assess the impact of modelling uncertainties on pulsar evolution by varying the birth spin period and surface magnetic field—using both fixed values and samples drawn from \citet{Faucher2006}—and by exploring the less constrained parameters of $\tau$ and $\kappa$ \citep{Chattopadhyay2020,Chattopadhyay2025}.
We adopt a fixed NS radius of $R_\mathrm{NS}=12\,\mathrm{km}$ and an inclination angle between the spin and magnetic axes of $\alpha=30^\circ$ \citep[as in][, who also showed that these choices have little effect on the results]{Chattopadhyay2020}, and take all remaining parameters directly from \texttt{MESA}.

\section{Results} \label{sec:results}
We follow the time evolution of the all 21 systems involving $\sim$1\,M$_\odot$ companion stars. We track their stellar properties, orbital changes, and mass-transfer, and examine the resulting effects on the spin and magnetic field of the NSs. The systems are then followed to their final stage as double compact object binaries, with the aim of drawing connections, where possible, to the observed Galactic MSP–WD population.

\subsection{Evolutionary Fate with \texttt{MESA}} \label{sec:futureMESA}

In all 21 binaries, the companion star fills its RL within $\sim10^9$–$10^{10}$\,yr.\footnote{Here we assume evolution from the zero-age main sequence; the time to RL overflow relative to the present epoch is obtained by subtracting the observed stellar age $t_{1,o}$.} Across all simulations, mass-transfer begins on the red giant branch (Case B) and proceeds to the end of either the red giant branch (all eccentric and some circularized runs) or the asymptotic giant branch (other circularized runs). Only in binary ID 16 of \texttt{cir-Ed} does a second stable RLOF episode that commences on the asymptotic giant branch (Case C), after Case B, while binary ID 15 of \texttt{cir-Ed} undergoes only Case C. Although the precise details vary, the 21 binaries in each of the four model simulations (Table~\ref{Tab:modelvariants}) exhibit broadly similar evolutionary pathways across a wide range of metallicities, initial stellar and orbital parameters. 
The subsequent phase of stable mass-transfer lasts $\sim10^5$–$10^6$\,yr in most of the eccentric runs and up to $\sim10^8$\,yr in the circularized runs. Longer phases of mass-transfer enables the NS to accrete sufficient material and gain angular momentum, significantly spinning it up and reducing its magnetic field. Following the mass-transfer episode(s), the companion evolves through the planetary nebula phase and becomes a WD. In eccentric models, only helium WDs form, whereas some circularized runs produce carbon–oxygen WDs (see Section~\ref{sec:WD}). Those that produce carbon–oxygen WDs exhibit thermal loops with pulses, which may be detectable by Rubin Observatory (see Section~\ref{sec:hr}). 

Although all of our models yield a recycled NS and a WD, the final orbital parameters, the amount of mass accreted by the NS, the WD composition, and the tracks in the HR diagram are quite different qualitatively between eccentric Eddington-limited, eccentric super Eddington and circular mass-transfer assumptions, producing divergent outcomes for the ensemble of 21 systems. A key qualitative distinction is the behaviour of the periastron separation: in eccentric prescriptions $r_{\rm p}$ remains approximately conserved through RLOF, whereas enforced circularization drives $r_{\rm p}$ upward with the secular expansion of $a$. For \texttt{cir-Ed}, we enforce artificial circularization at the onset of RLOF by setting $e=0$ throughout, following the standard assumption of efficient tidal dissipation \citep[e.g.][]{Hut1981}, consistent with the common prescription of instantaneous circularization adopted in literature and in binary evolution codes (e.g. \citealt{Hurley2002, Tauris2006}). This enforced circularization removes phase-dependent mass transfer at periastron and thereby suppresses the eccentricity pumping present in our eccentric prescriptions. This difference dictates much of the subsequent evolution we report below.

We show the time evolution, starting at the onset of RLOF ($t-t_{\rm RLOF}=0$, defined when the mass‐transfer rate $\dot{M}=10^{-10}$\,M$_\odot$\,yr$^{-1}$), of orbital eccentricity ($e$), orbital period ($P\mathrm{orb}$), periastron separation ($r_\mathrm{p}$), stellar radius ($R_1$), component masses ($M_1, M_\mathrm{NS}$), mass‐transfer rate ($\dot{M}$) and core masses ($M_\mathrm{core}$). We illustrate this with representative two systems, binary IDs 0 and 5\footnote{Selected as: (0) solar-metallicity system forming a helium WD; (5) solar-metallicity system forming a carbon-oxygen WD under circular mass-transfer.}, eccentric mass-transfer at the Eddington limit (\texttt{e-Ed}), and circularized mass-transfer at the Eddington limit (\texttt{cir-Ed}) in Fig.~\ref{fig:timeSeries}.

All three models--- \texttt{e-Ed}, \texttt{e-SpEd100}, and \texttt{e-SpEdcn}--- exhibit orbital expansion with rising eccentricity, sharper and higher mass-transfer peaks, and more abrupt stellar contraction. This behaviour is broadly consistent with the eccentric mass-transfer cycles reported by \citet{Rocha2025}, who likewise found enhanced and often intermittent accretion episodes in eccentric binaries across much of the parameter space they explored.
Indeed, the increasing eccentricity in all of our models shortly after the onset of MT is a direct consequence of the low initial mass ratio $q_\mathrm{i}$, which leads to positive $\dot{e}$ in the secular orbital evolution equations \citep{Sepinsky2007b}.
Additionally, in the context of a detailed stellar model and other competing secular effects such as tidal circularization, \citet{Rocha2025} showed that both $q_i$ and the initial eccentricity $e_\mathrm{i}$ of CO-hosting binaries (at the moment of RLOF) determine which systems remain eccentric post-MT via the relation:
\begin{align}
    e_\mathrm{crit} \simeq& -0.985 + 0.976 \, q - 0.1647 \, q^2 + 0.009232 \, q^3. 
    \label{eqn:q_e_polyfit}
\end{align}
The \texttt{cir-Ed} runs produce longer-lived, lower-amplitude transfer and smoother orbital growth. Despite these prescription-dependent differences, all channels follow the same broad sequence—rapid restructuring at RLOF followed by long-term decay—underscoring the robustness of the overall binary evolution pathway. As might be expected, the \texttt{e-SpEd100} model is intermediate to \texttt{e-Ed} and \texttt{e-SpEdcn} (the fully conservative case reaches $\sim\!3000\times$ the Eddington limit), but resembles \texttt{e-Ed} more closely. Accordingly, until the pulsar-evolution results (Section~\ref{sec:wd_pulsars}) we present only \texttt{e-SpEdcn} and \texttt{e-Ed} to illustrate the maximum variation.

In detail, conservation of angular momentum dictates a modest jump in orbital period for eccentric runs ($\Delta\log_{10} P_{\rm orb} \sim 0.2\text{--}0.6$), consistent with analytical expectations for non-conservative transfer \citep{Paczynski1967}. Roche-lobe stripping reduces the stellar radius by more than two orders of magnitude within $\sim10^5$\,yr, on its thermal time-scale \citep{Tauris2006}. The companion  mass plunges from $\simeq1$\,M$_\odot$ to $\simeq0.3$\,M$_\odot$ (e.g. Binary ID 0), while the NS accretes only a few $10^{-2}$\,M$_\odot$ in eccentric cases but more for circularized runs, owing to its higher effective accretion efficiency at the Eddington limit \citep{Shapiro1983}. This redistribution generates a transient $\dot M$ peak of $10^{-5}$–$10^{-4}$\,M$_\odot$\,yr$^{-1}$: \texttt{e-SpEdcn} yields the sharpest, earliest spike; \texttt{e-Ed} is intermediate; and \texttt{cir-Ed} sustains a broader plateau. Meanwhile, the helium core ($\sim0.35$\,M$_\odot$) and nascent carbon core ($\ll0.02$\,M$_\odot$) remain inert on RLOF time-scales. We refer to Figure.~\ref{fig:timeSeries} in the subsequent sections in more detail when necessary. The NSs in \texttt{e-SpEd100} and \texttt{e-SpEdcn} models accrete $\sim3$–$13\times$ and $\sim37$–$125\times$ more mass than those of \texttt{cir-Ed} and \texttt{e-Ed} cases, respectively.
Because $r_{\rm p}=a(1-e)$, the increase in $e$ during eccentric mass transfer offsets the modest growth of $a$, leaving $r_{\rm p}$ nearly unchanged; with $e=0$ in \texttt{cir-Ed}, $r_{\rm p}$ simply follows the expansion of $a$ and increases markedly.

\begin{figure*}[!htbp]
  \centering
  \begin{tabular}{@{}cc@{}} 
    \begin{minipage}[t]{0.5\textwidth}
      \includegraphics[width=\linewidth]{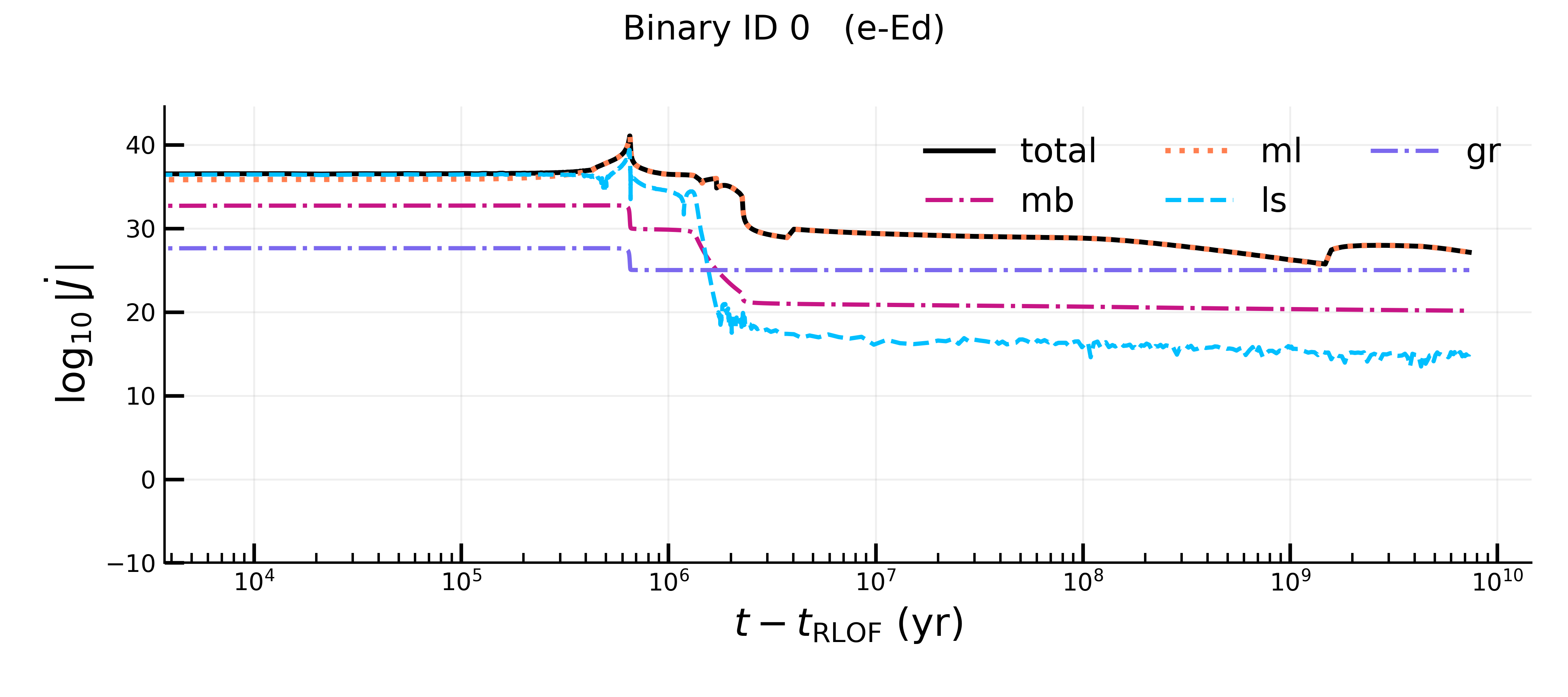}
    \end{minipage} &
    \begin{minipage}[t]{0.5\textwidth}
      \includegraphics[width=\linewidth]{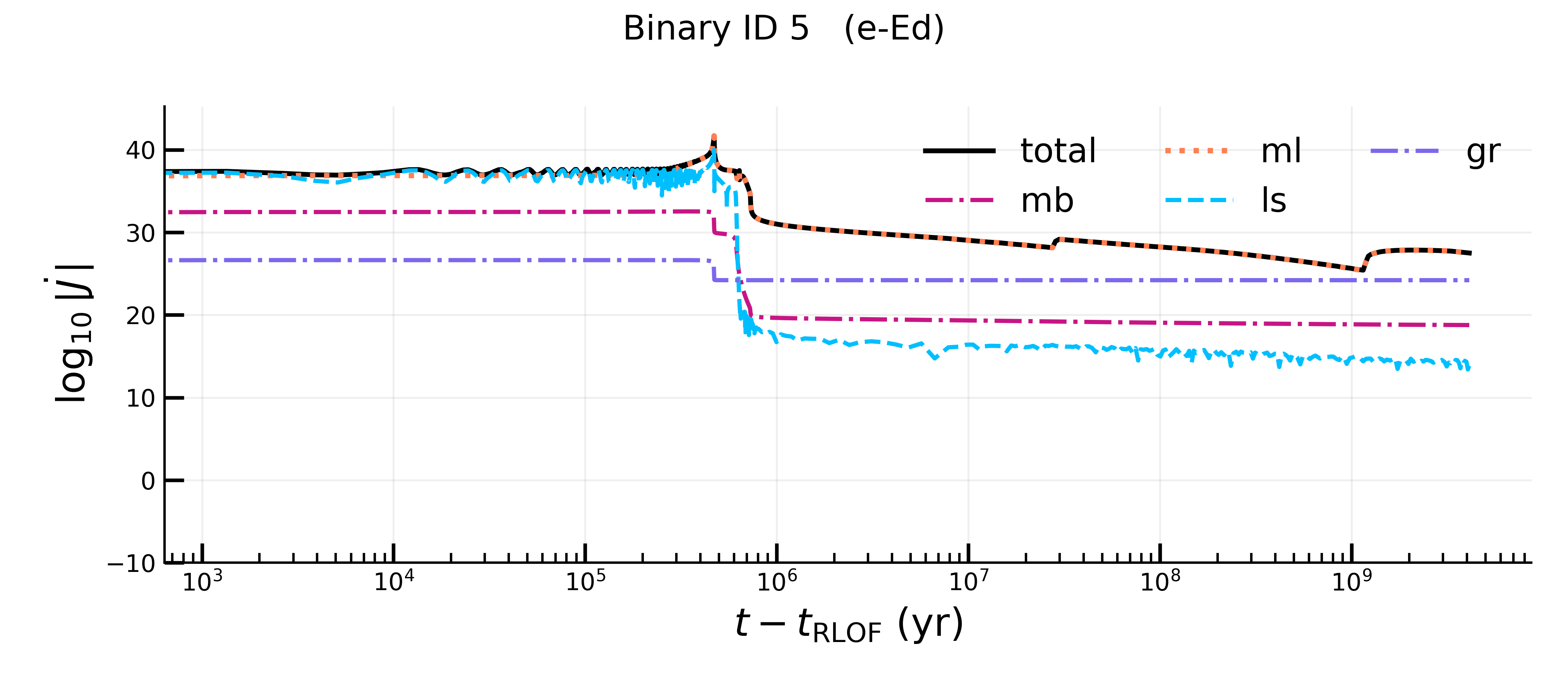}
    \end{minipage} \\
    \begin{minipage}[t]{0.5\textwidth}
      \includegraphics[width=\linewidth]{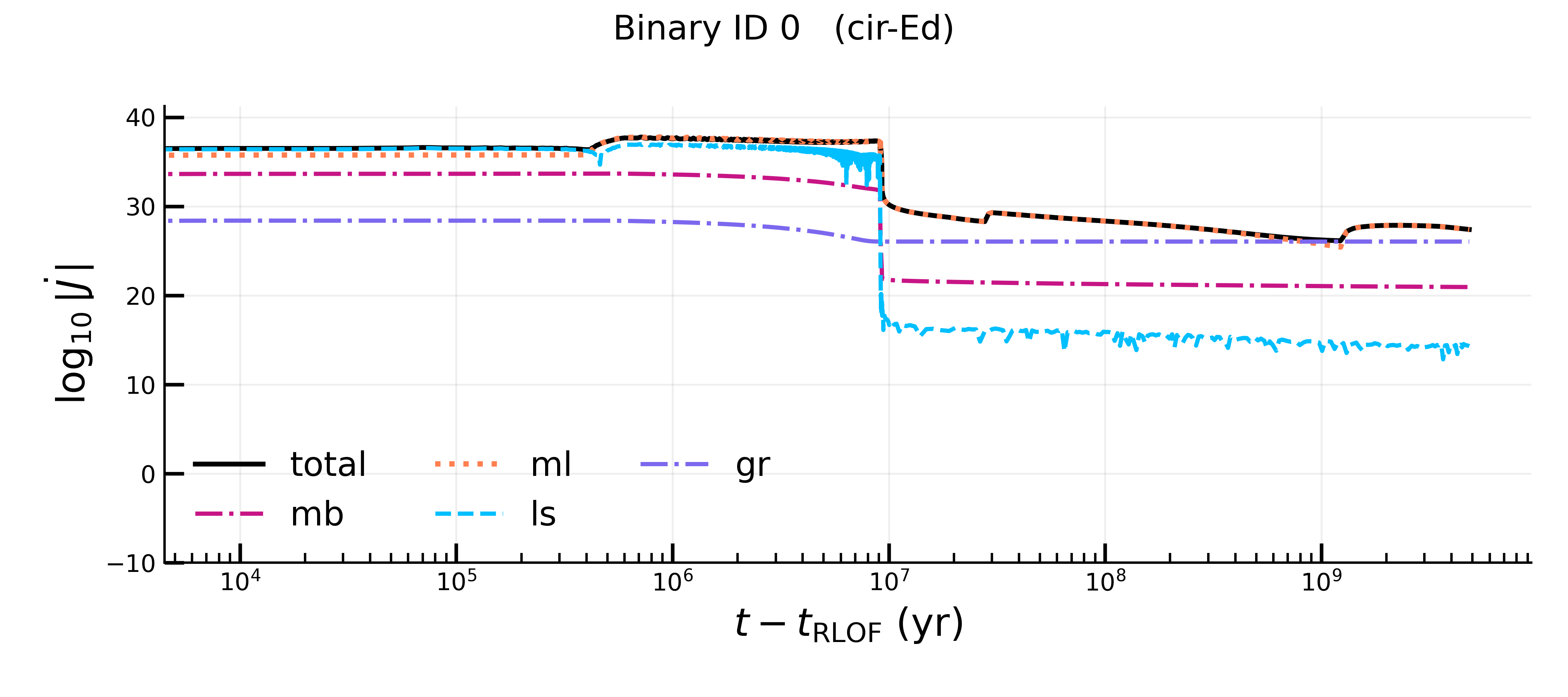}
    \end{minipage} &
    \begin{minipage}[t]{0.5\textwidth}
      \includegraphics[width=\linewidth]{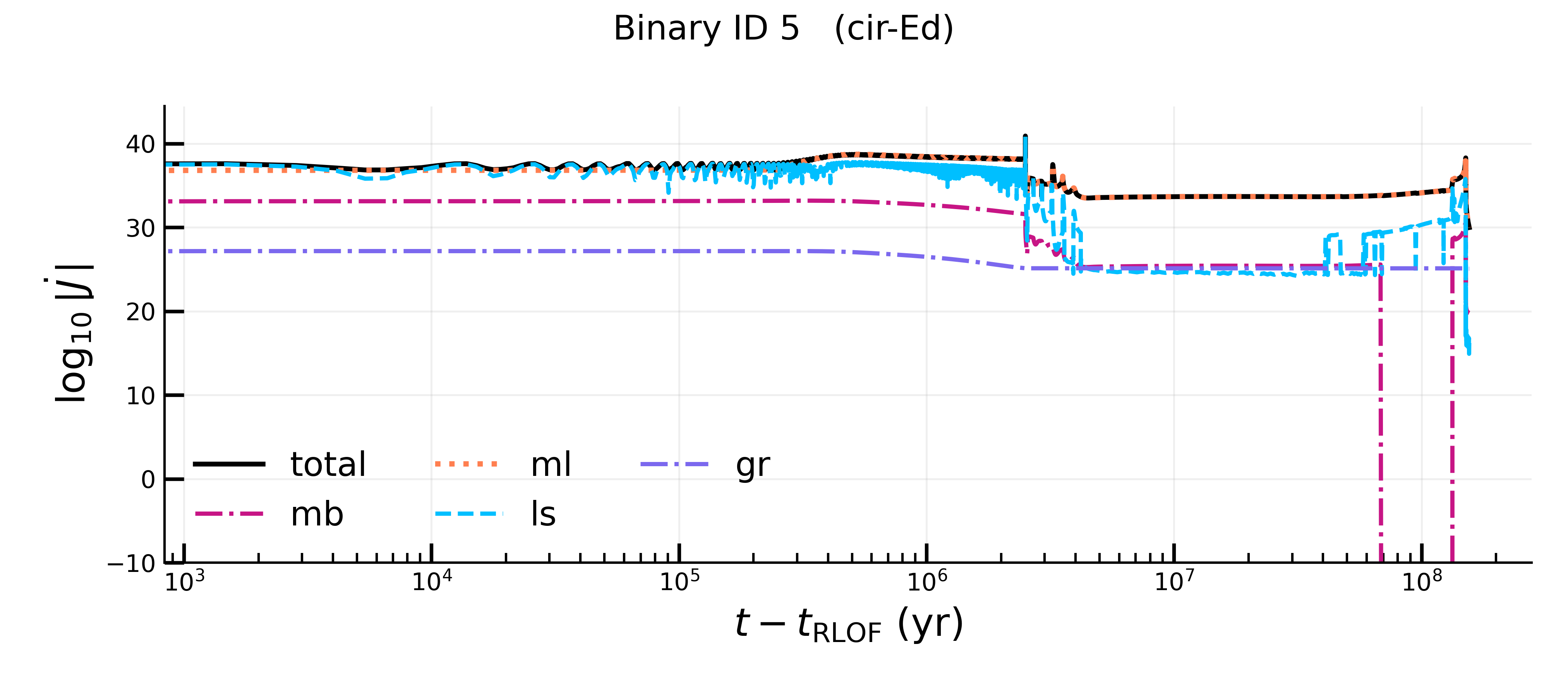}
    \end{minipage} \\
  \end{tabular}
  \caption{Time evolution of the angular‐momentum loss rate, $\log_{10}|\dot J|$, as a function of time since Roche‐lobe overflow, $t - t_{\rm RLOF}$. Top row shows eccentric runs \texttt{\detokenize{e-–Ed}} for binary models 0 (left) and 5 (right); bottom row shows the corresponding circular runs \texttt{\detokenize{cir-–Ed}}. In each panel the solid black curve is the total $|\dot J|$, while the patterned coloured lines denote the individual loss channels: mass loss (ml, orange dotted), magnetic braking (mb, magenta dashed), spin–orbit coupling (ls, cyan dash‐dotted), and gravitational radiation (gr, purple dash‐dot‐dot), as indicated in the legend. The orbital evolution is dominated by mass exchange and tides (at RLOF) while magnetic braking and gravitational radiation are subdominant.}
  \label{fig:jdotTimeSeries}
\end{figure*}

Figure~\ref{fig:jdotTimeSeries} shows the angular-momentum loss, decomposed into mass loss (`ml'), gravitational radiation (`gr'), magnetic braking (`mb'), and spin orbit (tidal) coupling (`ls'). At $t-t_{\rm RLOF}=0$, residual winds and tides set $\log_{10}|\dot J|\sim36$–37; magnetic braking contributes still negligibly at $\sim10^{32}$; gravitational radiation remains negligible. At RLOF, abrupt mass loss drives a short-lived spike to $\sim4\times10^{41}$\,erg\,s$^{-1}$, after which the tidal torque collapses, braking weakens with the shrinking convective envelope, and $|\dot J|$ decays over $10^6$–$10^{10}$\,yr, asymptoting to $\log_{10}|\dot J|\approx27$–28. At late times, gravitational wave emission becomes the dominant angular-momentum loss channel, as winds and braking fade away. Eccentric models show small tidal oscillations before the spike, reflecting periastron passages \citep{Sepinsky2007b}, while \texttt{cir-Ed} produces a smoother, broader angular-momentum peak: enforced circularization moderates the tidal response and sustains stronger braking.

\subsubsection{Orbital period-eccentricity evolution}\label{sec:porb_e}
\begin{figure}
    \centering
    \hspace{-1cm}
    \includegraphics[
  width=1.1\linewidth,
  height=6.1cm,
]{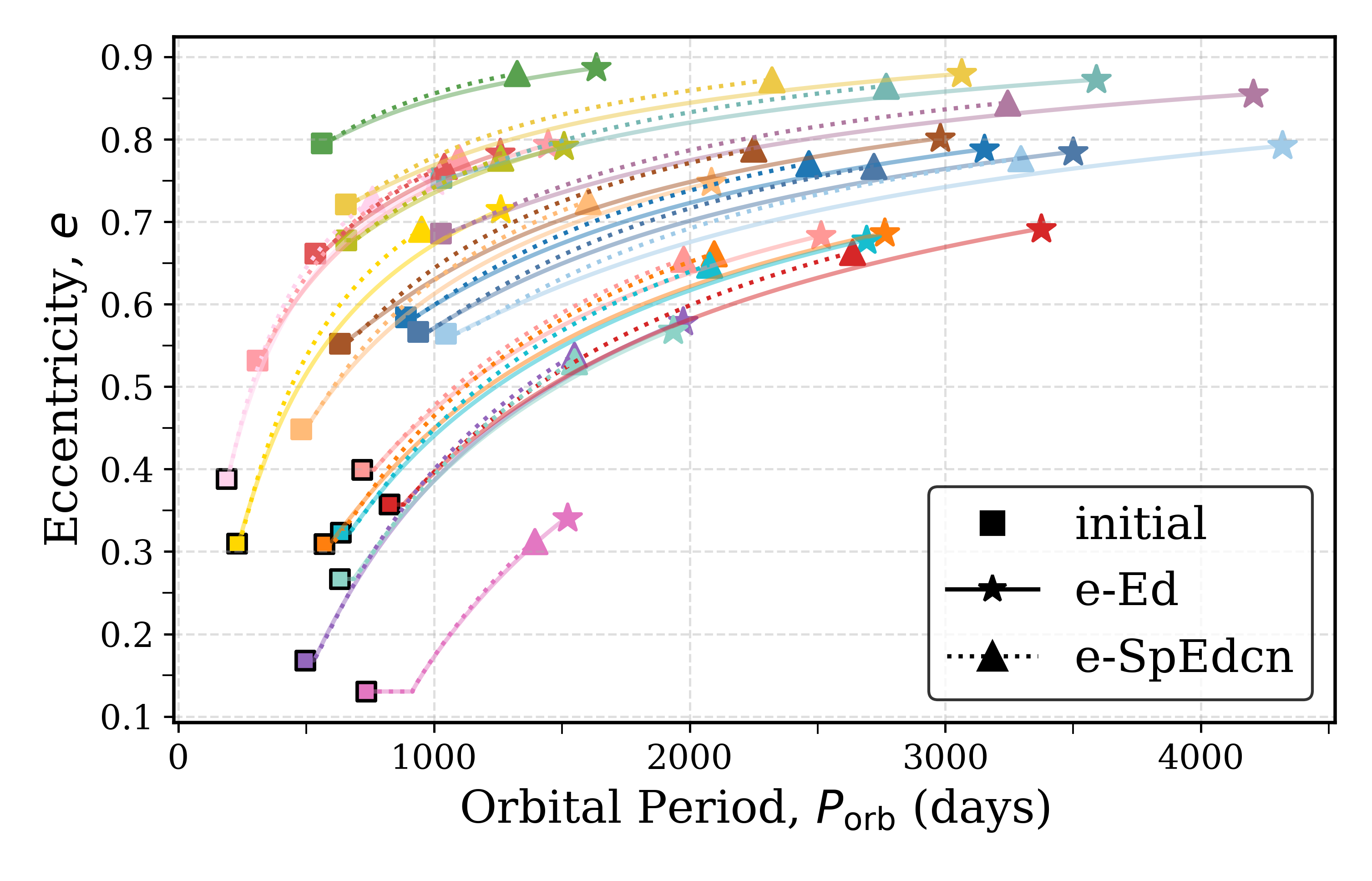}
\caption{Evolution of orbital period $P_\mathrm{orb}$ versus eccentricity $e$ for 21 Gaia NS binaries, starting from their observed values (square markers). We compare models with Eddington‐limited mass transfer (dashed curves, star markers, \texttt{e-Ed}) and fully conservative super‐Eddington mass transfer (dot–dashed curves, triangle markers, \texttt{e-SpEdcn}). Black‐edged squares denote systems that begin with $e \leq 0.4$, where the assumption of delta-function mass loss at periastron \citep{Sepinsky2007b} is less certain. In all cases, the systems expand and become more eccentric; however, the \texttt{e-SpEdcn} models show weaker expansion and end with slightly lower eccentricities.}
\label{fig:PorbE}
\end{figure}
When modelling the full case of eccentric mass-transfer, we are able to track the complete future evolution of orbital period $P_\mathrm{orb}$ versus eccentricity $e$ for the 21 Gaia NS binaries, as illustrated in Fig.~\ref{fig:PorbE} for models \texttt{e-Ed} and \texttt{e-SpEdcn}. In both cases, the orbits expand, but the \texttt{e-SpEdcn} binaries terminate at systematically shorter periods and lower final eccentricities. 
The orbit expands by a factor of about 2–5 for \texttt{e-Ed}, but only by a factor of $\sim$1.5–4 for \texttt{e-SpEdcn} (i.e. final periods systematically 20–30\% shorter than in \texttt{e-Ed}).
The \texttt{e-SpEdcn} systems also exhibit marginally lower final eccentricities ($\lesssim$10\%). Note that all systems, except one (binary ID 15, which starts with the least eccentric orbit and one of the widest separation), end at high eccentricities ($\sim$0.5–0.9).

In the \texttt{e-Ed} runs, only a small fraction of the donor's envelope is accreted during a brief $\sim10^{5}$–$10^{6}\,$yr single contact episode. The remainder of the mass‐transfer rate that exceeds the NS’s Eddington limit, $\dot{M}_{\rm loss}\;=\;\max\bigl(\dot{M}_0 - \dot{M}_{\rm Edd},\,0\bigr),$ is expelled from the system, carrying away the accretor’s specific orbital angular momentum \citep{Sepinsky2009}:
\begin{equation}
  \dot{J}_{\rm ml} = \dot{M}_{\rm loss}\;j_{\rm acc},\,\,
  j_{\rm acc} = \frac{M_{\rm NS}}{M_{\rm tot}}\,a^{2}\,\omega_{\rm orb}.
\end{equation}
where \(M_{\rm tot}=M_{\rm NS}+M_1\) the total mass, \(a\) the semi‐major axis, and \(\omega_{\rm orb}\) the instantaneous orbital angular frequency.  Because the orbital semi‐major axis depends sensitively on the total orbital angular momentum \(J_{\rm orb}\) via
\begin{equation}
  a \;\propto\; J_{\rm orb}^{2}\,M_{\rm tot}^{-1}\,\mu^{-3}, 
  \qquad \mu \equiv \frac{M_{\rm NS}\,M_1}{M_{\rm tot}},
\end{equation}
and as \(P_\mathrm{orb}\propto a^{3/2}\), the extra angular‐momentum loss in the Eddington‐limited case induces a larger fractional contraction of both \(a\) and \(P_{\rm orb}\) than in the fully conservative, super‐Eddington models.  Quantitatively, we find that
\begin{equation}
  \frac{P_{\rm orb, Edd}}{P_{\mathrm{orb,}\beta=1}}
  \;\simeq\;
  \Bigl(\frac{a_{\rm Edd}}{a_{\beta=1}}\Bigr)^{3/2}
  \;\approx\; 0.75\text{–}0.84,
\end{equation}
in excellent agreement with the numerical results of \citet{Sepinsky2009}.   
Overall, these NSs accrete only $\sim10^{-2}$\,M$_{\odot}$ of companion material before detachment, leaving the NSs only as mildly recycled pulsars.

In the \texttt{e-SpEdcn} runs; nearly the entire transferred envelope ($\sim0.5$–$1.0\,M_{\odot}$) is retained over a roughly similar $\sim10^{5}$–$10^{6}\,$yr interval. Because the envelope is removed more quickly, the companion  radius $R_1$ shrinks on a shorter time-scale than the RL, leading to earlier ($\sim10^{4}$–$10^{5}\,$yr than \texttt{e-Ed}) detachment. These \texttt{e-SpEdcn} binaries therefore accrete far more mass but reach detachment at marginally smaller $P_{\rm orb}$ and lower $e$ than their Eddington‐limited counterparts. This reduced orbital expansion aligns closely with the numerical findings of \citet{Sepinsky2009}. The \texttt{e-SpEd100} NSs, however, gain about 0.1\,M$_\odot$ (roughly same as \texttt{cir-Ed}, Figure~\ref{fig:timeSeries}).

In contrast to eccentric cases \texttt{cir-Ed} models undergo steady, thermal–timescale mass transfer of order \(10^{6}\)–\(10^{7}\,\mathrm{yr}\) and end with \(P_{\rm orb}\sim250\)–\(1200\)\,days (\(\approx0.5\)–\(3\times\) the initial; Fig.~\ref{fig:PorbE}). Although circularization at RLOF can initially shrink the orbit, the subsequent phase generally widens it as the companion mass \(M_1\) decreases. Most \texttt{cir-Ed} runs therefore expand to \(1.5\)–\(3\times\) the initial period, while systems starting with large eccentricities (\(e\sim0.6\)–0.8) lose enough orbital energy during circularization to finish shorter (binary IDs 4, 11, 13, 14). The outcome reflects the competition between circularization-driven energy loss and mass-transfer-driven widening.

Following the \citet{Kolb1990} prescription, \texttt{cir-Ed} treats RLOF as continuous throughout the orbit. With \(q=M_1/M_{\rm NS}\approx0.7<1\), the orbit widens as \(a\) increases, the Roche lobe grows, and \(\dot M_1\) declines smoothly. In \texttt{e-Ed}, only a fraction is accreted by the \(\sim1.4\,\mathrm{M_\odot}\) NS and the remainder is expelled with the accretor’s specific angular momentum—tempering but not reversing the secular expansion. In practice, the NS gains \(\sim3\times10^{-2}\)–\(2\times10^{-1}\,\mathrm{M_\odot}\), with most cases clustering near \(0.1\,\mathrm{M_\odot}\). 


\begin{figure}
    \centering
\hspace{-1cm}
    \includegraphics[
  width=1.1\linewidth,
  height=6.1cm,
]{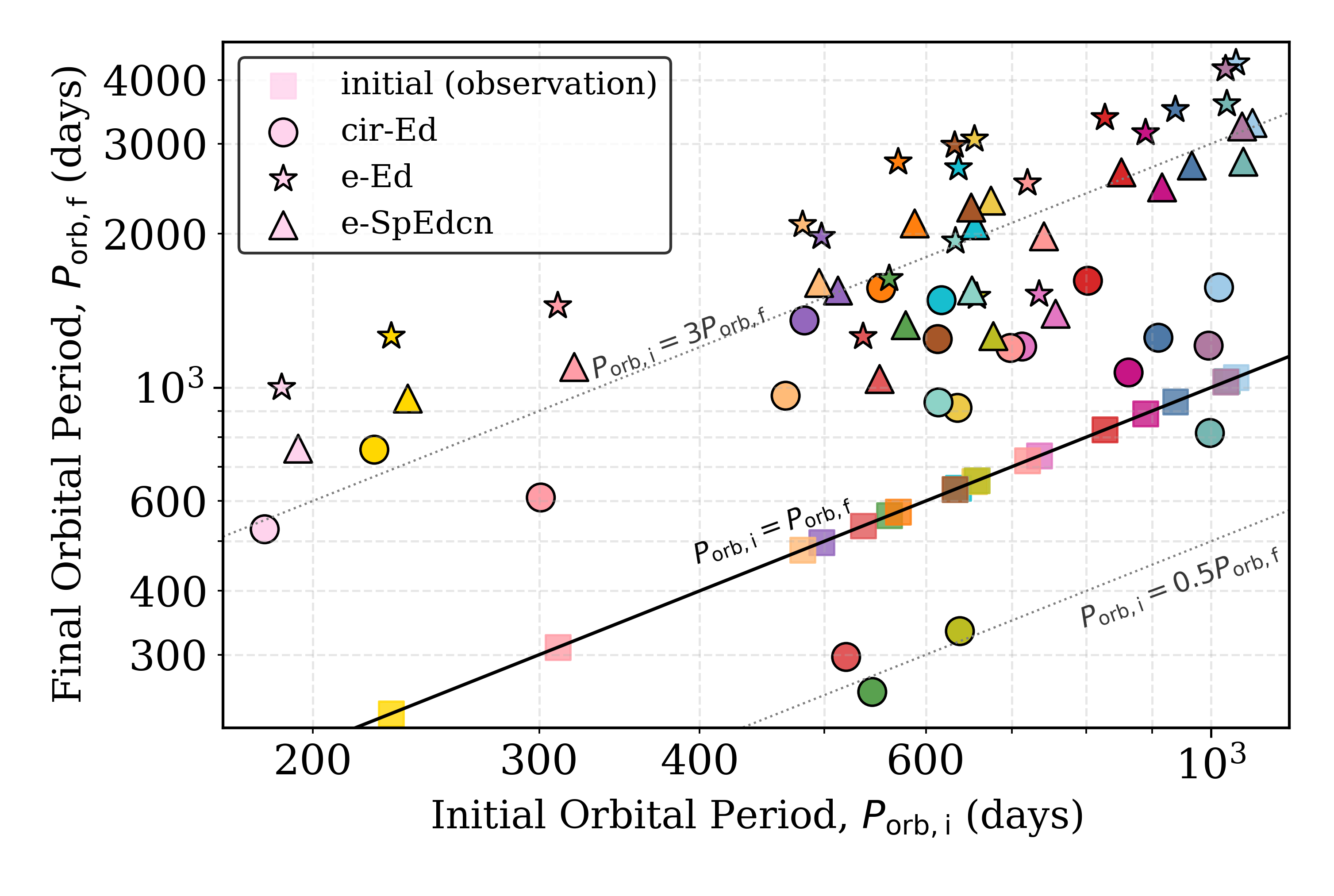}
    \caption{Comparison of initial (square) and final orbital periods:
\texttt{\detokenize{cir-Ed}} (circle), \texttt{\detokenize{e-Ed}} (star),
\texttt{\detokenize{e-SpEdcn}} (triangle). Coloured squares mark the initial
(observed) periods; colours are consistent per 21 individual systems. The solid black guide is
\(P_\mathrm{orb,1}=P_\mathrm{orb,f}\) (no net change). Dotted gray guides are \(P_\mathrm{orb,1}=0.5\times P_\mathrm{orb,f}\) and \(P_\mathrm{orb,1}=3\times P_\mathrm{orb,f}\) (indicating a factor-of-two decrease and a threefold increase, respectively). Most final orbits of the binaries increases after full evolution (stable mass transfer and formation of the WD), but four systems in \texttt{\detokenize{cir-Ed}} end with periods shorter than their initial values. These binaries expand during stable mass transfer but not enough to recover the orbital period lost during enforced circularization at RLOF.}
    \label{fig:PorbAll}
\end{figure}

\subsubsection{White Dwarfs}\label{sec:WD}
\begin{figure}
    \centering
    \includegraphics[
  width=1.1\linewidth,
  height=6.1cm,
]{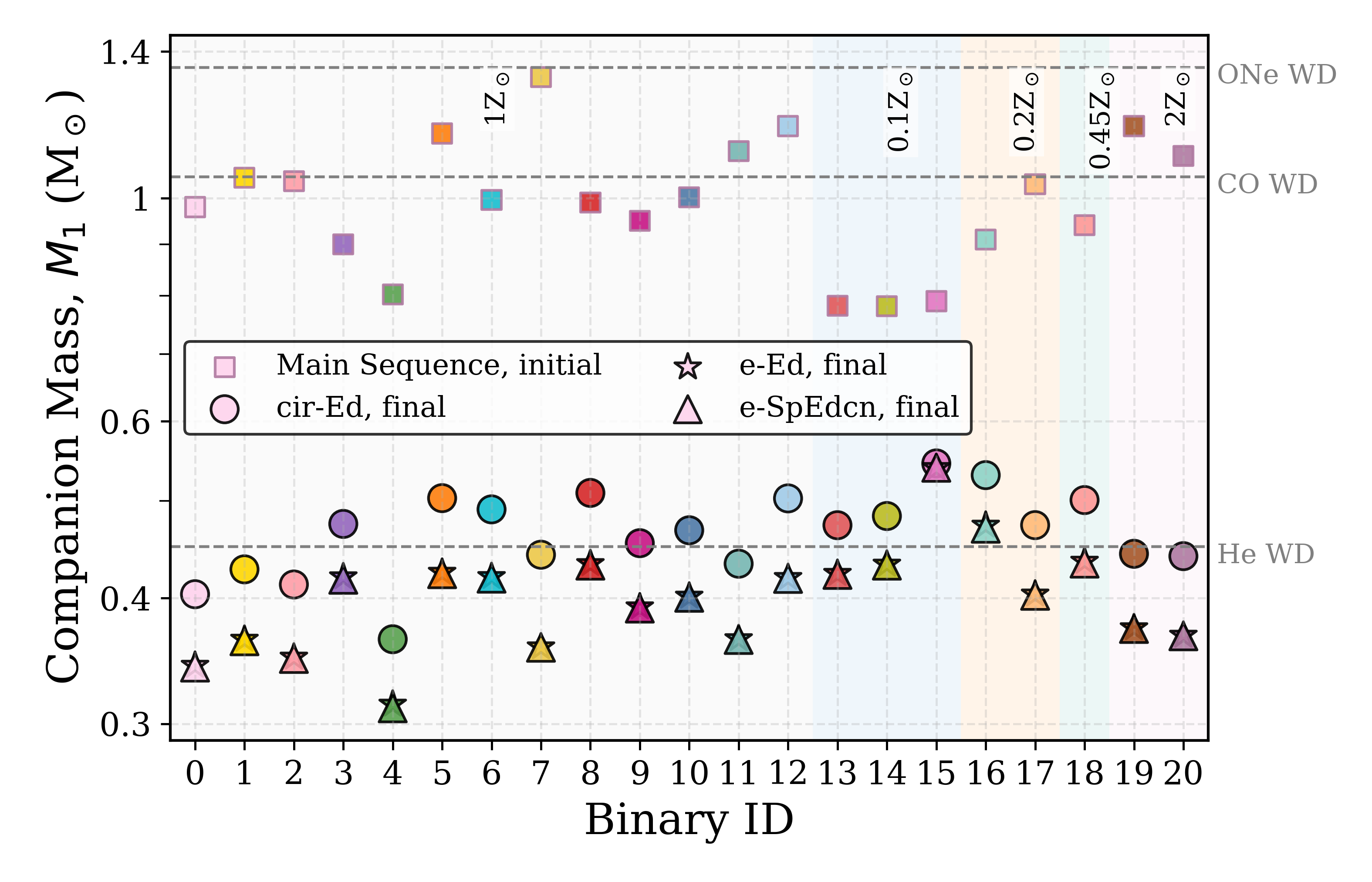}
\caption{Companion mass $M_1$ across 21 Gaia NS binary calculations. Squares denote observed (initial) donor masses; stars, triangles, and circles show the final masses from eccentric Eddington-limited (\texttt{\detokenize{e-Ed}}), fully conservative super-Eddington (\texttt{\detokenize{e-SpEdcn}}), and circularized Eddington-limited at RLOF (\texttt{\detokenize{cir-Ed}}) models, respectively. Shaded vertical bands indicate the metallicity grid ($Z/Z_\odot$). Horizontal dashed lines mark nominal upper mass boundaries for He, CO, and ONe WDs. Although a few eccentric cases (IDs 15–16) lie marginally above the He–CO transition line, their final profiles and central abundances confirm helium‐core compositions. These global mass cutoffs can bias WD-type inference for near-threshold systems, so we instead rely on core-mass fractions and full stellar profiles, which reveal that eccentric-transfer remnants are consistently He WDs, while only wide circularized systems cross into the CO regime}

\label{fig:WDall}
\end{figure}

All our models end with the evolving star becoming a WD, defined as the epoch when the central Coulomb coupling parameter reaches $\Gamma_{\rm c}=20$ \citep{Fragos2023}, thereby forming a NS–WD binary.
In each simulation, we classify the remnant white dwarf by examining the final \texttt{MESA} output for its core composition.  Specifically, we compute the total mass contained in the helium core ($M_{\rm He}$), the carbon core ($M_{\rm C}$), and the oxygen core ($M_{\rm O}$) and form the fractions
\[
  f_{\rm He} \;=\;\frac{M_{\rm He}}{M_{\rm core}}, \\
  f_{C+O} \;=\;\frac{M_{\rm C} + M_{\rm O}}{M_{\rm core}},
\]
where, $M_{\rm core}=M_{\rm He}+M_{\rm C}+M_{\rm O}$.
If $f_{\rm He}\ge0.90$, the remnant is classified as a helium white dwarf (He WD); if $f_{C+O}\ge0.90$ with both $M_{\rm C}/M_{\rm core}>0.10$ and $M_{\rm O}/M_{\rm core}>0.10$, it is classified as a carbon–oxygen white dwarf (CO WD); if instead $f_{C+O}\ge0.90$ and $M_{\rm O}/M_{\rm core}\ge0.60$, it is classified as an oxygen–neon white dwarf (ONe WD) \citep{Driebe1998,Fontaine2001,Renedo2010,Istrate2014}. This bulk‐fraction approach ensures that the classification reflects the full evolutionary history of the star, which is more robust than relying on a single final profile that may be ambiguous in transitional cases or affected by numerical gradients. In the rare marginal cases not resolved by bulk fractions, we classify by central abundances (ONe WD if $X_{\rm O}+X_{\rm Ne}\ge0.95$ and $X_{\rm Ne}\ge0.05$; as CO WD if $X_{\rm C}+X_{\rm O}\ge0.95$; or as He WD if $X_{\rm He}\ge0.95$) to ensure a robust classification. The final masses of the NS and the WD are shown in Figure~\ref{fig:WDall}, and the final masses as well as the WD-type are shown in Table~\ref{tab:finalvaluesNSWD}. 

Applying this procedure to all eccentric mass‐transfer runs—\texttt{e-Ed}, \texttt{e-SpEd100} and \texttt{e-SpEdcn}—reveals that every such run ends with $f_{\rm He}\gtrsim0.95$ and negligible C–O burning. Hence, all eccentric mass‐transfer remnants are He WDs.

For \texttt{cir-Ed}, only those models (nine of 21) whose periastron‐equivalent circular periods $P_{\rm orb}$ (Table~\ref{Tab:GaiaNSBinaries}) are wide enough for the helium core to exceed the ignition mass $M_{\rm He,crit}\simeq0.45$\,M$_{\odot}$ before envelope stripping ultimately produce CO WDs. This ignition threshold corresponds to the well‐established core mass at which helium ignites under degeneracy in low–mass stars \citep{deBoer2017,Vos2020,Irrgang2020,Laplace2025}. It is a robust result of stellar evolution; modern models, including \texttt{MESA}, consistently yield, $M_{\rm He,crit}=0.45$–$0.47$\,M$_\odot$, with only minor variations ($\lesssim0.02$\,M$_\odot$) arising from metallicity, convective mixing, or neutrino cooling \citep{Arancibia2024,Rodrig2025PASA}. Binary IDs 3, 5, 6, 8, 10, and 12 initiate RLOF as Case-B donors with $P_{\rm orb}\approx300$–400\,days (binary ID 10 already in core-He burning), while system 15 is a pure Case-C donor ($P_{\rm orb}\approx700$\,days) and system 16 follows a hybrid Case-B$\rightarrow$C path from $P_{\rm orb}\approx400$\,days. In these wider binaries, the donors remain in stable RLOF for $\sim10^{6}$–$10^{7}$\,yr, allowing their helium cores to grow beyond $M_{\rm He,crit}$ before the envelope is exhausted; they thus end as CO WDs with $M_{\rm WD}\gtrsim0.47$\,M$_\odot$. The remaining \texttt{cir-Ed} models—those starting at shorter periods or with lower donor masses—detach earlier with $M_{\rm core}<M_{\rm He,crit}$ and therefore produce He WDs. This dichotomy underscores the key difference from eccentric runs, where periastron-confined mass transfer always terminates too early for helium ignition.


\subsubsection{Hertzsprung Russell Diagram}\label{sec:hr}

\begin{figure*}[!htbp]
  \centering
  \begin{tabular}{@{}cc@{}} 
    \begin{minipage}[t]{0.5\textwidth}
      \includegraphics[width=\linewidth]{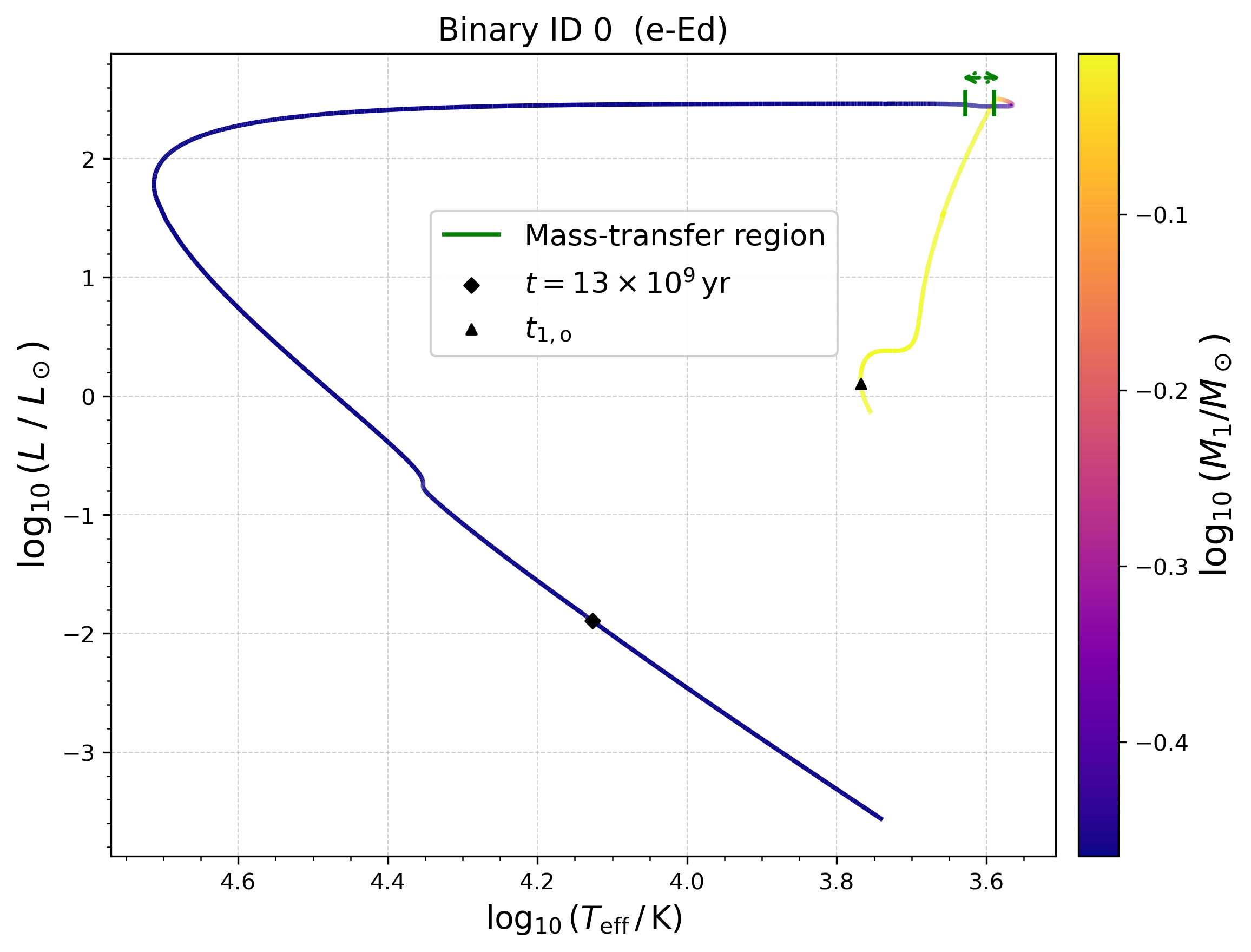}
    \end{minipage} &
    \begin{minipage}[t]{0.5\textwidth}
      \includegraphics[width=\linewidth]{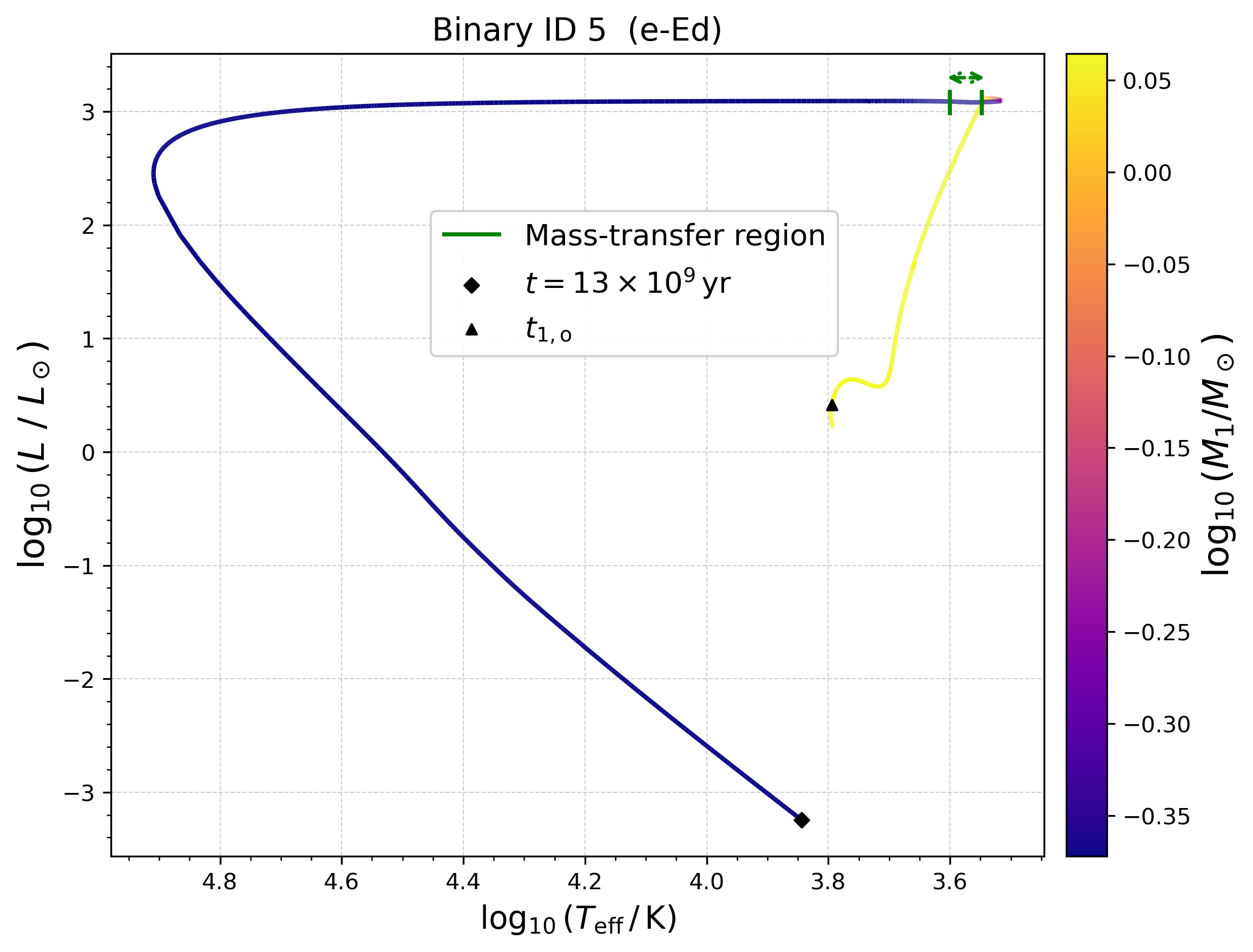}
    \end{minipage} \\
    \begin{minipage}[t]{0.5\textwidth}
      \includegraphics[width=\linewidth]{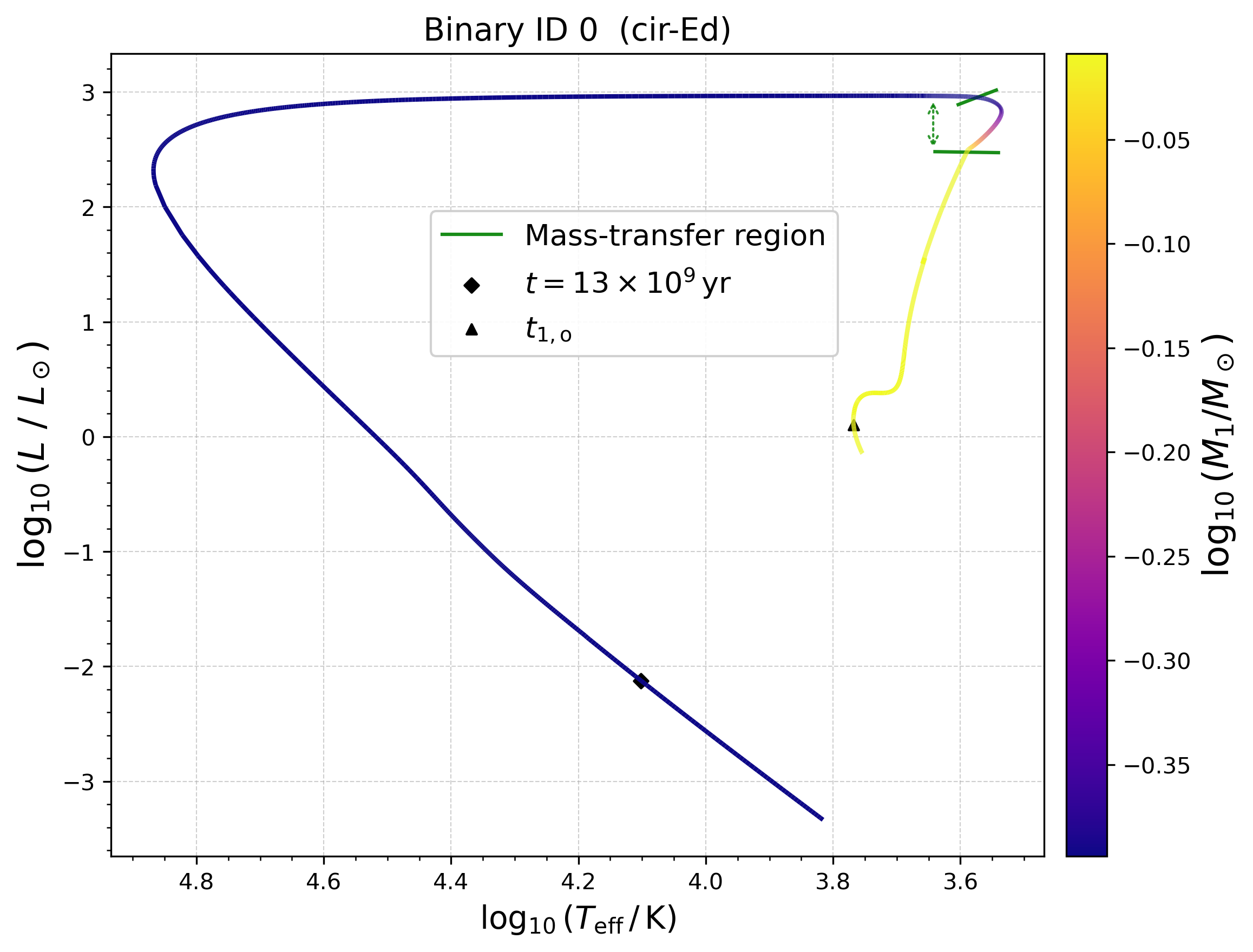}
    \end{minipage} &
    \begin{minipage}[t]{0.5\textwidth}
      \includegraphics[width=\linewidth]{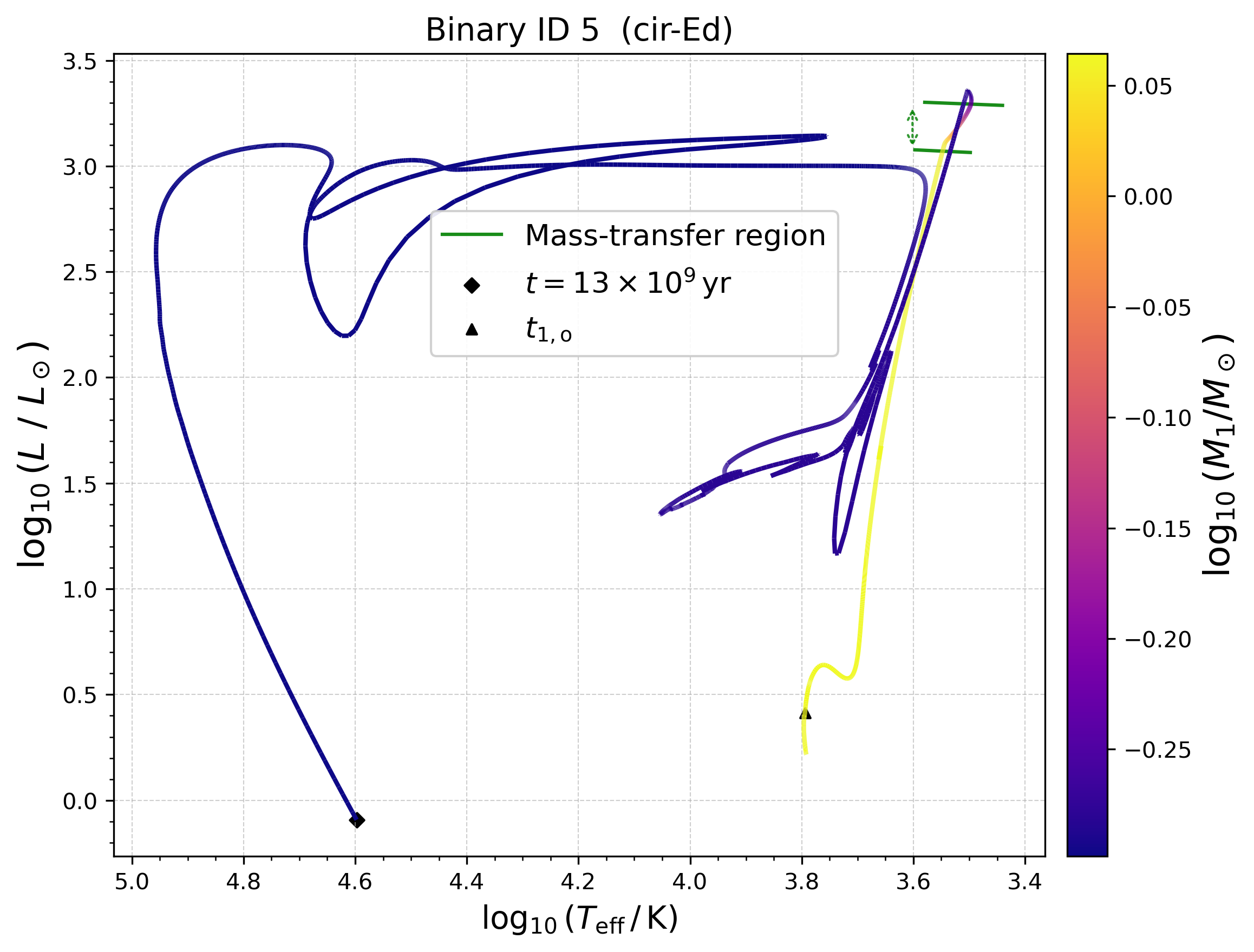}
    \end{minipage} \\
  \end{tabular}
  \caption{Hertzsprung–Russell diagrams showing the evolution of effective temperature  
($T_{\rm eff}$) versus luminosity ($L$) for our early Case B Eddington‐limited  
mass‐transfer runs in eccentric orbits (\texttt{\detokenize{e-–Ed}}; top row) and circular orbits  
(\texttt{\detokenize{cir-–Ed}} ; bottom row).  Each panel displays binary systems 0 (left) and 5  (right), with the track coloured by instantaneous stellar mass, log$_{10}(M/M_\odot)$  
(see colourbars).  Phases of active mass-transfer 
are highlighted by green circles: for model 0 the \texttt{\detokenize{cir-–Ed}} green points form a continuous band at the tip of the giant‐branch plateau, whereas in the \texttt{\detokenize{e-–Ed}} run they cluster more tightly (mass shedding only at periastron). In the bottom‐right panel, model 5 of the circular case, the donor undergoes pronounced thermal loops before settling as a carbon–oxygen white dwarf. The black diamond marks $t = 13\times10^9\,$yr,  
and the black triangle denotes the Gaia‐inferred shift epoch, $t_{\rm shift}$.  
Horizontal axis is log$_{10}(T_{\rm eff}/{\rm K})$ (decreasing to the right),  
vertical axis is log$_{10}(L/L_\odot)$.}
\label{fig:HRdiag}
\end{figure*}

We discuss the Hertzsprung–Russell (HR) evolutionary tracks for \texttt{e-Ed}, \texttt{e-SpEdcn}, and \texttt{cir-Ed}. The diagrams illustrate how the donor’s thermal and structural response to mass transfer determines whether it ends as a He or CO WD. In eccentric systems, envelope removal occurs rapidly, preventing the occurrence of hydrogen-shell flashes \citep[e.g.][]{Althaus2001, Istrate2014} or late-time pulses \citep{Lawlor2023}. 
In these models, the donor detaches directly from the red-giant branch as a He WD with no loop signature in its HR track (see Fig.~\ref{fig:timeSeries}, \ref{fig:HRdiag}). In \texttt{e-SpEdcn}, fully conservative transfer occasionally produces a brief luminosity–temperature adjustment loop \citep{Tauris2006}, but the core still detaches marginally below ignition, again leaving a faint He WD.

In contrast, \texttt{cir-Ed} ensures that the envelope loss proceeds continuously on the donor’s thermal time scale. Here, the He core can continue to grow during the red-giant plateau phase; if it surpasses the ignition mass $M_{\rm He,crit}$ (Sec.~\ref{sec:WD}), central helium ignition drives the characteristic blueward knee and clockwise loop in the HR diagram \citep{Webbink1975, Siess2007, Driebe1998, Nelson2004}. Models crossing this threshold detach with $M_{\rm core}\gtrsim0.47$\,M$_\odot$ and evolve as CO WDs, while those that fail to ignite the He core resemble the eccentric \texttt{e-Ed} tracks and end as He WDs. Finally, late thermal pulses (LTPs) may occur after AGB departure, producing short-lived HR loops at ionizing temperatures. 

\begin{table*}[ht]
  \centering
  \begin{tabular}{l 
                  cc    
                  ccc   
                  ccc   
                  ccc   
                  cc}   
    \toprule
      & \multicolumn{2}{c}{$e$}
      & \multicolumn{3}{c}{$P_\mathrm{orb}$ (days)}
      & \multicolumn{3}{c}{$M_\mathrm{NS}$ (M$_\odot$)}
      & \multicolumn{3}{c}{$M_\mathrm{WD}$ (M$_\odot$)}
      & \multicolumn{2}{c}{WD‐type} \\
    ID
      & \texttt{e-Ed} & \texttt{e-SpEdcn}
      & \texttt{e-Ed} & \texttt{e-SpEdcn} & \texttt{cir-Ed}
      & \texttt{e-Ed} & \texttt{e-SpEdcn} & \texttt{cir-Ed}
      & \texttt{e-Ed} & \texttt{e-SpEdcn} & \texttt{cir-Ed}
      & \texttt{e-Ed} & \texttt{cir-Ed} \\
    \hline
    0  & 0.75 & 0.73 & 1000.07 &  759.24 &  528.06 & 1.339 & 1.961 & 1.506 & 0.343 & 0.341 & 0.404 &   He    &    He   \\
    1  & 0.71 & 0.69 & 1260.38 &  951.57 &  755.80 & 1.318 & 1.985 & 1.434 & 0.364 & 0.362 & 0.427 &   He    &   He    \\
    2  & 0.79 & 0.78 & 1445.10 & 1096.97 &  609.23 & 1.331 & 2.007 & 1.481 & 0.349 & 0.348 & 0.413 &   He    &   He    \\
    3  & 0.58 & 0.54 & 1974.92 & 1548.09 & 1352.99 & 1.290 & 1.735 & 1.334 & 0.420 & 0.417 & 0.474 &   He    &   CO    \\
    4  & 0.89 & 0.88 & 1633.77 & 1324.53 &  253.66 & 1.353 & 1.825 & 1.632 & 0.314 & 0.311 & 0.364 &    He   &   He    \\
    5  & 0.69 & 0.66 & 2763.28 & 2094.98 & 1565.88 & 1.368 & 2.078 & 1.413 & 0.424 & 0.423 & 0.503 &   He    &   CO    \\
    6  & 0.68 & 0.65 & 2691.35 & 2077.80 & 1482.32 & 1.404 & 1.949 & 1.453 & 0.420 & 0.418 & 0.490 &   He    &    CO   \\
    7  & 0.88 & 0.87 & 3063.71 & 2321.84 &  912.45 & 1.707 & 2.656 & 1.860 & 0.358 & 0.357 & 0.442 &   He    &  He     \\
    8  & 0.69 & 0.66 & 3374.36 & 2636.99 & 1618.07 & 1.527 & 2.050 & 1.563 & 0.433 & 0.431 & 0.509 &   He    &   CO    \\
    9  & 0.79 & 0.77 & 3151.54 & 2464.55 & 1070.07 & 1.404 & 1.937 & 1.478 & 0.392 & 0.391 & 0.454 &    He   &   He    \\
    10 & 0.79 & 0.77 & 3499.25 & 2719.54 & 1251.61 & 1.450 & 2.026 & 1.517 & 0.402 & 0.400 & 0.467 &   He    &   He    \\
    11 & 0.87 & 0.87 & 3590.65 & 2767.00 &  814.82 & 1.408 & 2.142 & 1.530 & 0.364 & 0.364 & 0.433 &  He     &   He    \\
    12 & 0.79 & 0.78 & 4317.94 & 3294.58 & 1570.05 & 1.486 & 2.222 & 1.538 & 0.419 & 0.418 & 0.503 &   He    &   CO    \\
    13 & 0.78 & 0.77 & 1258.01 & 1039.13 &  296.96 & 1.302 & 1.636 & 1.401 & 0.423 & 0.422 & 0.473 &   He    &  He     \\
    14 & 0.79 & 0.78 & 1508.02 & 1260.82 &  333.60 & 1.391 & 1.713 & 1.478 & 0.433 & 0.431 & 0.483 &   He    &   He    \\
    15 & 0.34 & 0.31 &  1522.57 &  1394.45 & 1202.95 & 1.898 & 2.033 & 1.898 & 0.539 & 0.539 & 0.545 &   He    &   CO    \\
    16 & 0.57 & 0.53 & 1934.80 & 1548.78 &  935.26 & 1.268 & 1.664 & 1.279 & 0.473 & 0.469 & 0.530 &   He    &    CO   \\
    17 & 0.75 & 0.73 & 2084.27 & 1602.34 &  963.69 & 1.480 & 2.088 & 1.558 & 0.404 & 0.402 & 0.473 &   He    &    CO   \\
    18 & 0.68 & 0.66 & 2513.56 & 1976.70 & 1195.27 & 1.370 & 1.842 & 1.413 & 0.435 & 0.433 & 0.501 &   He    &    CO   \\
    19 & 0.80 & 0.79 & 2979.26 & 2250.50 & 1244.04 & 1.396 & 2.185 & 1.485 & 0.374 & 0.373 & 0.443 &    He   &    He   \\
    20 & 0.86 & 0.84 & 4204.31 & 3243.12 & 1207.79 & 1.611 & 2.328 & 1.712 & 0.367 & 0.367 & 0.440 &   He    &   He    \\
    \hline
  \end{tabular}
  \caption{Final‐state summary (eccentricity, period, and masses) for models \texttt{e-Ed} (quite similar to \texttt{e-SpEd100}), \texttt{e-SpEdcn} and \texttt{cir-Ed}. WD‐type columns are added for \texttt{e-Ed} (identical to \texttt{e-SpEdcn} and \texttt{e-SpEd100}) and \texttt{cir-Ed}.}
  \label{tab:finalvaluesNSWD}
\end{table*}

\subsection{Connection with pulsar-white dwarfs} \label{sec:wd_pulsars}

A central goal of this work is to assess whether the Gaia NS–MS binaries can evolve into systems resembling the Galactic MSP–WD population. The latter provides robust observational anchors—well-measured $P_\mathrm{spin}$, $\dot{P}_\mathrm{spin}$, and, in some cases, orbital properties to companion masses—all quantities that map directly onto our model outputs. Importantly, the MSP–WD sample spans the two contrasting accretion geometries we aim to explore: continuous, circular RLOF and eccentric, mostly periastron-limited transfer. Their observed demographics therefore provide a direct empirical benchmark for evaluating the consequences of each evolutionary pathway.

Field radio surveys (excluding globular clusters) have uncovered nearly two hundred MSP–WD systems, almost all in circular or very low-eccentricity orbits, consistent with long-lived, stable phases of RLOF \citep{Bhattacharya1991,Kiziltan2010}. At the same time, mounting evidence suggests that many ``canonical” NS–WDs may also arise through CE evolution, with both theoretical \citep{MacLeod2015} and observational \citep{Yang2025} indications of significant pulsar recycling during the CE. Thus, both CE and stable RLOF channels likely contribute to the observed MSP–WD population.

In this work, however, the Gaia progenitors are modelled exclusively through the latter pathway—stable RLOF under both eccentric and circular prescriptions. While CE recycling has been extensively examined for double-NS and NS–BH systems \citep{Chattopadhyay2020,Chattopadhyay2021}, a dedicated population synthesis for NS–WD binaries remains outstanding. Preliminary results from our parallel study \citep{Chattopadhyay2025inprep} likewise point to a significant role for unstable mass transfer in producing canonical MSP–WDs. By contrast, here we forward-evolve the 21 Gaia NS–MS binaries and find that RLOF remains stable across all prescriptions tested. This stability introduces a key degeneracy: under the standard assumption of circularization at RLOF onset, prolonged stable transfer yields MSPs in circular NS–WD binaries—indistinguishable from the dominant observed population.

\begin{figure}
    \centering
    \includegraphics[width=1\linewidth]{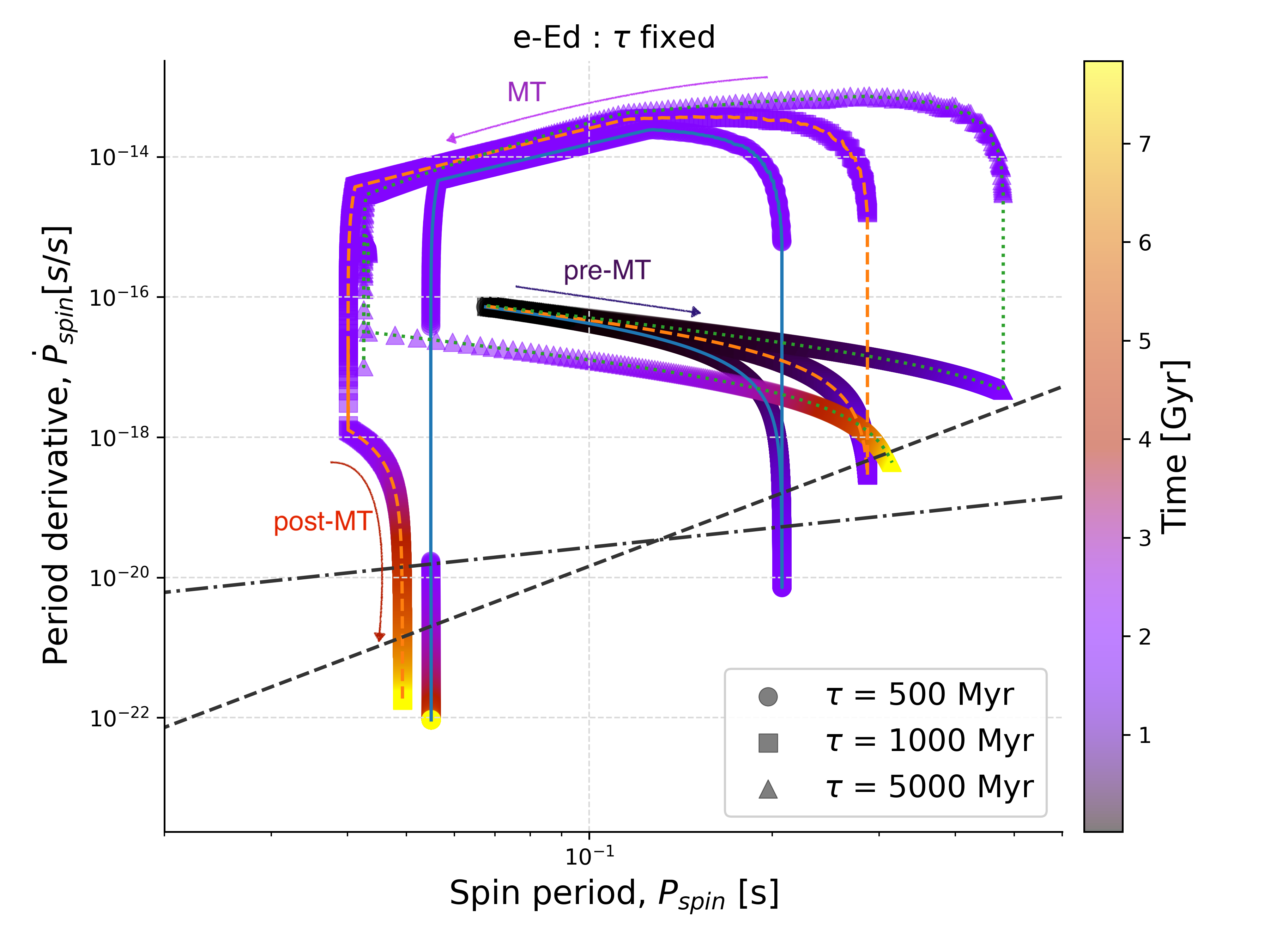}
    \includegraphics[width=1\linewidth]{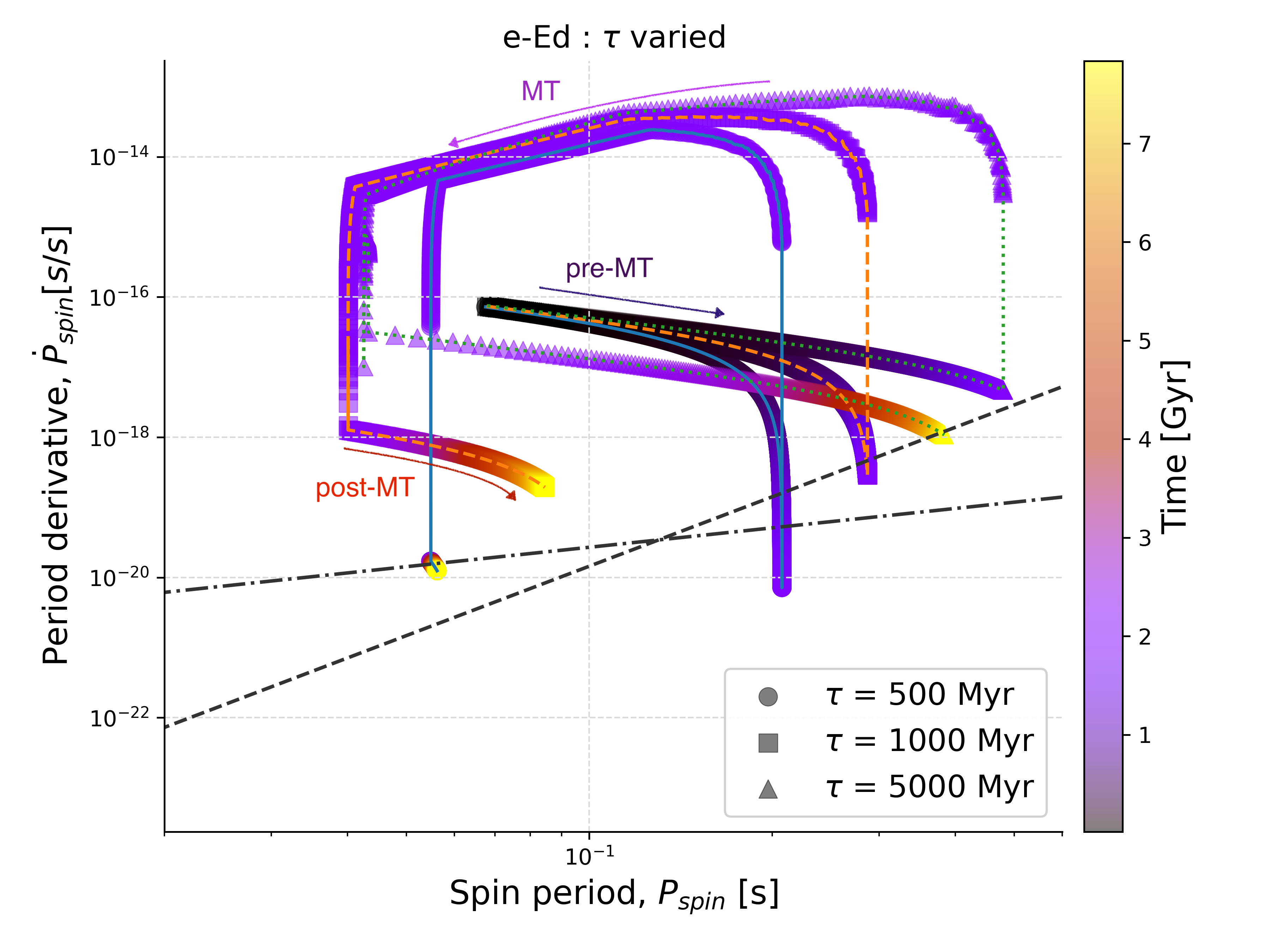}
\caption{Evolution of pulsar spin–down in the \texttt{\detokenize{e-–Ed}} model, showing $\dot P_{\rm spin}$ versus spin period $P_{\rm spin}$ coloured by system age.  Symbols mark the end of mass-transfer (MT) for three magnetic‐field decay time-scales, $\tau=500\,$Myr (circle), 1000\,Myr (square) and 5000\,Myr (triangle). Top: field decays with a fixed $\tau$; Bottom: $\tau\propto B_r^{-1}$.  The dashed and dotted–dashed lines show the death-lines. Allowing $\tau$ to scale inversely with $B$ (bottom) prolongs the radio-loud lifetime of recycled pulsars relative to fixed-$\tau$ models (top). Arrows denote the evolutionary phases--- pre-MT, MT, and post-MT. Notably, the post-MT phase is substantially extended under the adaptive $\tau$ prescription, consistent with the observed longevity of recycled radio pulsars.}
\label{fig:pulsartauComparison}
\end{figure}
\begin{figure*}[!htbp]
  \centering
    \includegraphics[width=\textwidth]{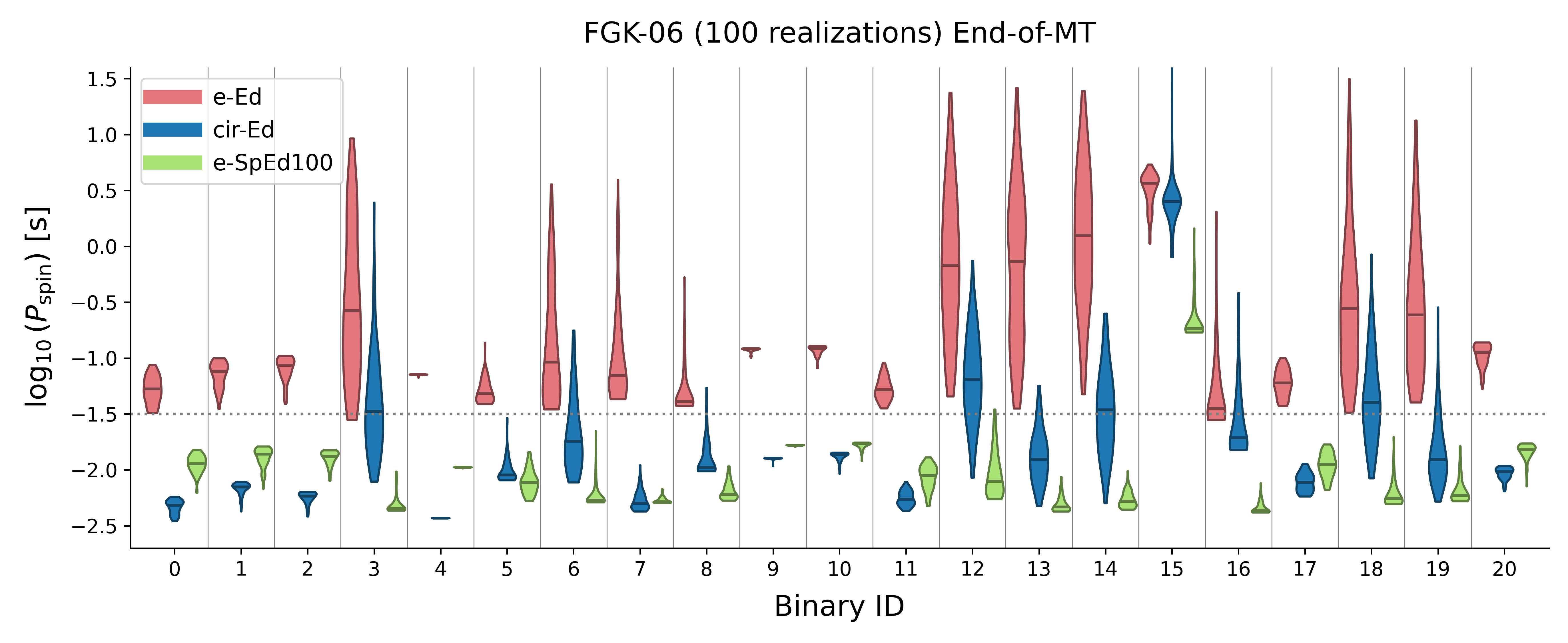}
    \includegraphics[width=\textwidth]{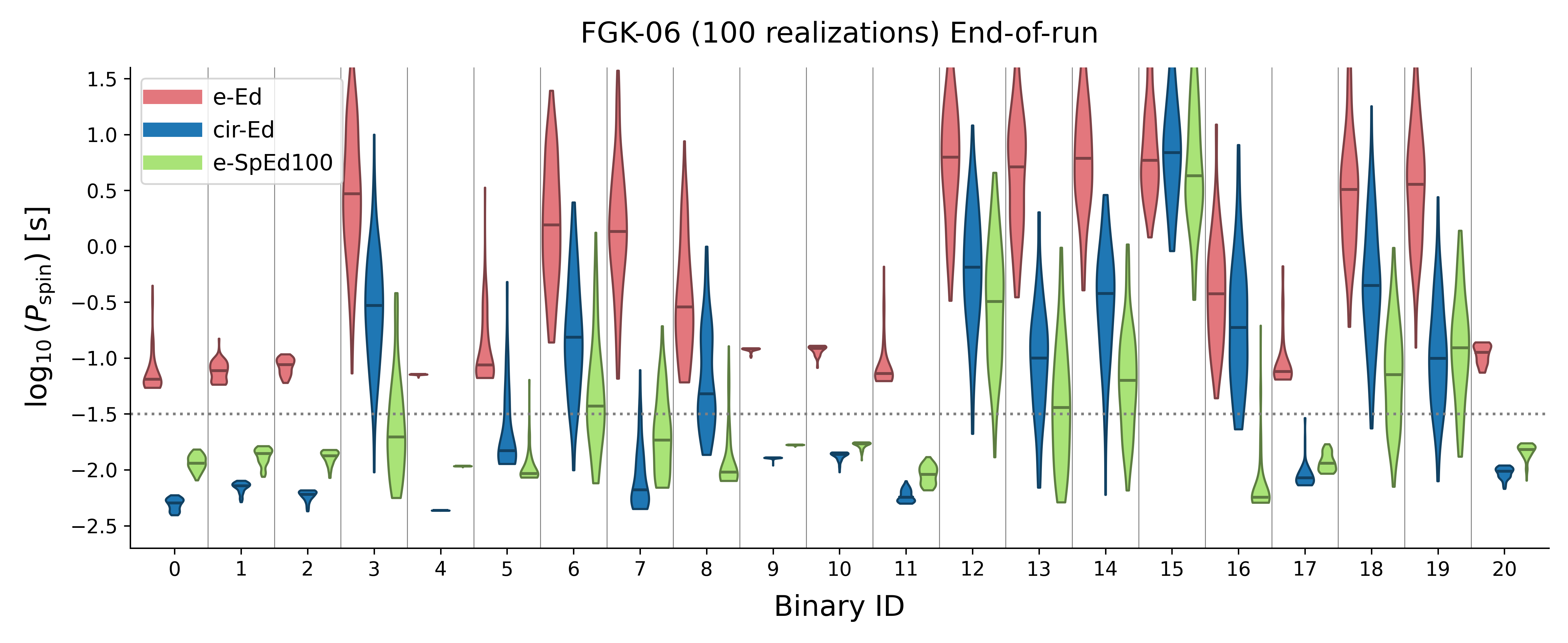}
\caption{
Violin distributions of $\log_{10}P_\mathrm{spin}$ for the 21 Gaia binaries, each with 100 Monte Carlo realizations of initial NS spin and surface magnetic field drawn from the distributions of \citet{Faucher2006} (FGK–06). Shown are results for eccentric Eddington-limited (\texttt{e-Ed}, red), circular Eddington-limited (\texttt{cir-Ed}, blue), and eccentric super-Eddington (\texttt{e-SpEd100}, green) models.
Top: end of mass transfer (MT). Bottom: end of run, after subsequent spin-down to WD formation. 
The thick horizontal bar indicates the median; the dotted line marks the MSP regime ($P_{\rm spin}\le 30\,\mathrm{ms}$). 
\texttt{cir-Ed} models overwhelmingly produce MSPs at the end of MT (16/21 systems have median $P_{\rm spin}<30\,\mathrm{ms}$), whereas \texttt{e-Ed} models typically peak at $0.1$–$1\,\mathrm{s}$ and rarely cross the MSP threshold. 
Super-Eddington \texttt{e-SpEd100} models also predominantly yield MSPs (16/21) at the end of MT. 
With our adaptive, surface-field–dependent field-decay timescale, systems with narrowly peaked end-of-MT spin distributions tend to remain MSPs for Gyr timescales, whereas systems with broader end-of-MT distributions often evolve out of the MSP regime by WD formation. This demonstrates that only circular or strongly super-Eddington eccentric transfer produces systems consistent with the observed MSP–WD population in terms of NS spin.
}
  \label{fig:FGK06}
\end{figure*}

\subsubsection{Spin and surface magnetic field evolution}\label{sec:pulsarTrack}
Each of the 21 Gaia systems’ NS is assigned an initial spin and magnetic field, which are evolved under equations~\ref{eqn:isolatedSpin} and~\ref{eqn:isolatedMagneticField} until the onset of mass transfer, and subsequently under equations~\ref{eqn:accSpin} and~\ref{eqn:accMagneticField} during accretion. After detachment, each pulsar is further evolved until WD formation.   

Representative $P_\mathrm{spin}$-$\dot{P}_\mathrm{spin}$ evolutionary tracks are shown in Figure~\ref{fig:pulsartauComparison}, marking the phases of pre-mass transfer, stable mass transfer, and post-mass transfer. These highlight the influence of different assumptions for the pulsar field-decay timescale $\tau$, as well as our adaptive model for $\tau$ (pulsar surface magnetic field decay mass-scale; Equation~\ref{equ:tau}). The adaptive prescription allows MSPs to remain radio-loud significantly longer than a fixed-$\tau$ model.  

Because the epoch at which recycled pulsars are observed post mass-transfer is uncertain, our comparisons adopt the pulsar state immediately after detachment, or at WD formation (end-of-run), unless otherwise noted.

Modelling pulsar evolution involves substantial uncertainties, in addition to those already present in binary evolution (and its associated mass transfer). Pulsar-specific sources of uncertainty include the initial spin period, the surface magnetic field strength, as well as the characteristic time-scale of magnetic field decay ($\tau$) and the associated mass-scale $\kappa$.

While the observed clustering of young pulsars at particular spin-down rates provides guidance (and thus an estimate of their surface magnetic fields), there is currently no accurate prescription to map pre-supernova stellar properties to a neutron star’s birth spin period and magnetic field strength. Even if such a prescription existed, our Gaia systems are already observed as neutron stars, so their birth pulsar properties cannot be reliably reconstructed. 

Since each of the NSs in our sample of 21 Gaia systems may have been born at any birth spin and magnetic field ($P_0, B_0$); within a certain range, we perform a suite of Monte Carlo variations for each systems. Fixing $\tau = 1000,$Myr and $\kappa = 0.025,$Myr (guided by the best matches to Galactic double-NS systems; \citealt{Chattopadhyay2020,Chattopadhyay2021,Sgalletta2023}), we draw $P_0$ and $B_0$ from the distributions of \citet{Faucher2006}:
\begin{align*}
  \log_{10}B_0 &\sim \mathcal{N}\bigl(12.65,\;0.55^2\bigr)\,,\\
  P_0 &\sim \mathcal{N}\bigl(0.30\,\mathrm{s},\;(0.15\,\mathrm{s})^2\bigr)\,, 
  \quad P_0>0\,.
\end{align*}

Each system is sampled 100 times over the birth-parameter distributions. 
Figure~\ref{fig:FGK06} summarises the post–mass-transfer (detachment) and end-of-run (WD formation) $P_\mathrm{spin}$ values for the three channels actually shown (\texttt{cir-Ed}, \texttt{e-Ed}, \texttt{e-SpEd100}). 
For completeness, we note that \texttt{e-SpEdcn} accretes $\sim3$–$13\times$ more mass than \texttt{cir-Ed} and $\sim37$–$125\times$ more than \texttt{e-Ed}; nearly all such runs yield MSPs, but they are omitted from Fig.~\ref{fig:FGK06} to avoid redundancy. 
At detachment, the circular models overwhelmingly yield MSPs (defined as $P_\mathrm{spin}\leq 30$\,ms), with nearly all realisations clustering at $\log_{10} P_\mathrm{spin}\lesssim -2$ ($P_\mathrm{spin}\lesssim 10$\,ms), showing a tight interquartile range and only a few long-period outliers.

The \texttt{e-Ed} systems typically peaks around $\log_{10} P_\mathrm{spin}\sim -1$ to $0$ and seldom reaches the millisecond regime. This dichotomy reflects the higher angular-momentum transfer efficiency of continuous, phase-averaged accretion in circular systems relative to the impulsive accretion in eccentric mass transfer. 

The \texttt{e-SpEd100} model, however, allows the NS to accrete at 100 times the Eddington limit, and thus produces MSPs in roughly half of the binaries. 

\begin{table}
\centering
\hspace{-2cm}
\begin{tabular}{c|cccc}
\hline
Model & $P_0$ [ms] & $B_0$ [G] & $\tau$ [Myr] & $\kappa$ [Myr] \\
\hline
A & 50  & $10^{11}$   & 1000 & 0.025 \\
B            & 150 & $10^{11}$  & 1000 & 0.025 \\
C            & 50  & $10^{12}$  & 1000 & 0.025 \\
D            & 150 & $10^{12}$  & 1000 & 0.025 \\
E            & 50  & $10^{11}$   & 1000 & 0.25  \\
F            & 50  & $10^{11}$   & 500  & 0.025 \\
G            & 50  & $10^{11}$   & 5000 & 0.025 \\
\hline
\end{tabular}
\caption{Summary of pulsar model variations (initial spin, magnetic fields and parameters surface magnetic field time-decay and mass-decay time-scale) around the fiducial case (A).}
\label{tab:modelVariationsPulsars}
\end{table}

\begin{figure}
    \centering
    \includegraphics[width=1\linewidth]{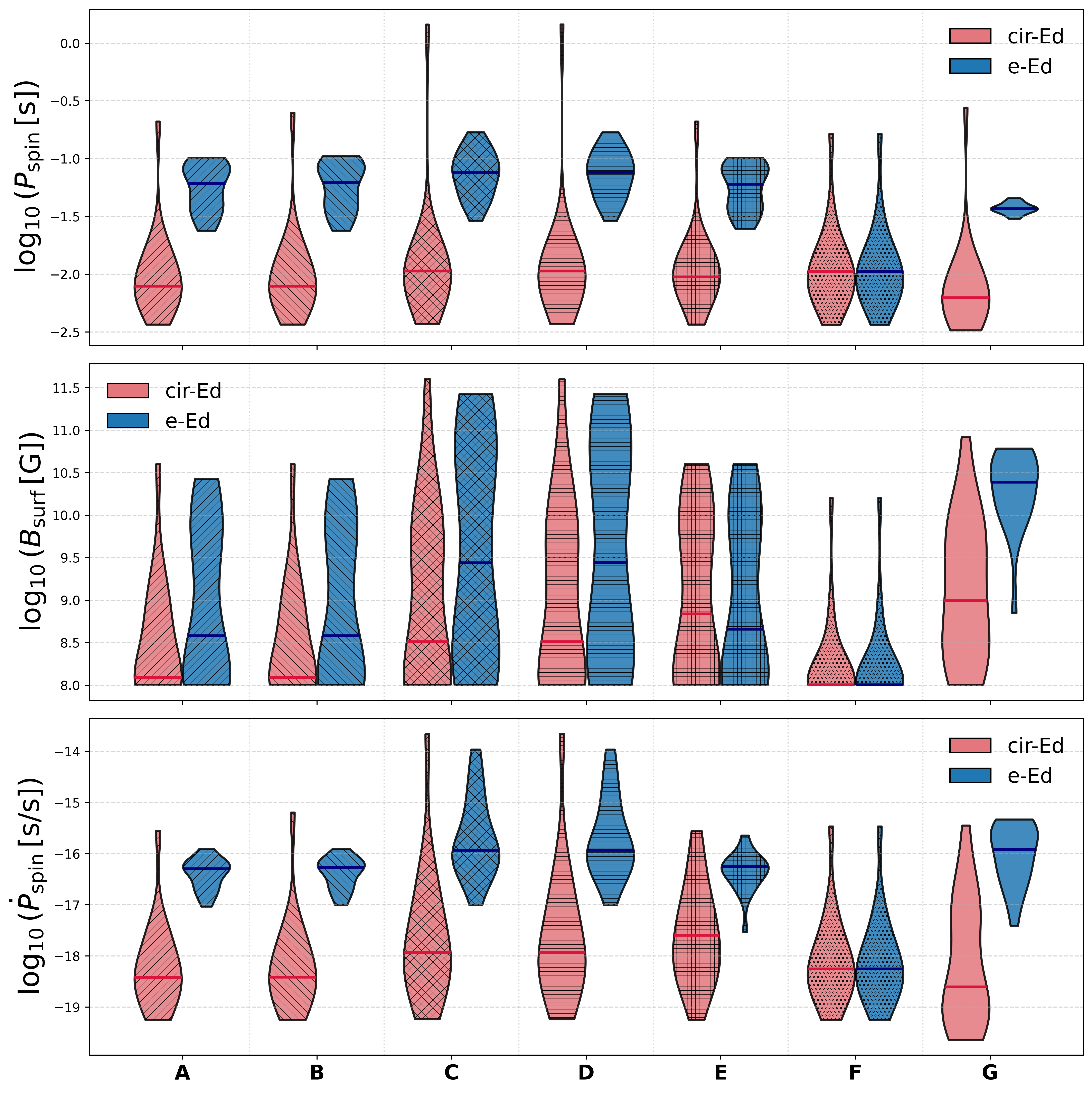}
\caption{Violin plots of final spin period (top), surface magnetic field (middle) and spin‐down rate (bottom) for all 21 Gaia binaries in pulsar models A–G (each model has a different fill), comparing circular (\texttt{\detokenize{cir-–Ed}}, pink) and eccentric (\texttt{\detokenize{e-–Ed}}, blue) mass-transfer. Each box spans the interquartile range (25th–75th percentiles), the middle bright line shows the median value, and whiskers show the full range of outcomes.  Models B–D vary the birth spin and surface magnetic fields, while E–G probe the impact of the magnetic field decay and mass decay time-scales. Irrespective of uncertainties in birth parameters or decay prescriptions, for Eddington-limited accretion, circular RLOF overwhelmingly yields highly recycled MSPs, whereas eccentric RLOF typically produces only mildly recycled pulsars, illustrating the robustness of our results despite of uncertainties in modelling pulsar evolution.}
\label{fig:pulsarABCD}
\end{figure}
While the FGK–06 prescriptions remain a common choice for pulsar birth parameters, recent studies (e.g.\ \citealt{Chattopadhyay2020}; \citealt{Sgalletta2023}) favour uniform distributions—or even fixed values for recycled pulsars—in both spin and magnetic field. Moreover, although the Monte Carlo sampling of each individual binary for FGK-06 explores system-level trends in MSP formation, it is equally important to examine whether the ensemble of 21 systems exhibits systematic differences under the eccentric (\texttt{e-Ed}) versus circularized (\texttt{cir-Ed}) mass-transfer prescriptions. Unlike the more indirect influence of $P_0$ and $B_0$, the pulsar evolutionary parameters $\tau$ and $\kappa$ directly govern the spin and magnetic-field evolution of recycled pulsars.  

To assess the sensitivity of our results to these uncertainties, we explore seven additional model variations around a fiducial case (A). In case (A), we adopt $P_0 = 50\,$ms, $B_0 = 10^{11}\,$G, $\tau = 1000\,$Myr, and $\kappa = 0.025\,$Myr \citep[guided by][]{Chattopadhyay2020}. Holding $\tau$ and $\kappa$ fixed, models B–D vary the birth parameters: (B) $P_0 = 150\,$ms, $B_0 = 10^{11}\,$G; (C) $P_0 = 50\,$ms, $B_0 = 10^{12}\,$G; and (D) $P_0 = 150\,$ms, $B_0 = 10^{12}\,$G. Finally, retaining the fiducial $P_0$ and $B_0$ of (A), models (E)–(G) adjust a single evolutionary timescale: (E) $\kappa = 0.25\,$Myr, (F) $\tau = 500\,$Myr, and (G) $\tau = 5000\,$Myr.  

This suite of scenarios, summarised in Table~\ref{tab:modelVariationsPulsars}, enables us to isolate the influence of both birth and evolutionary parameters on the final spin and magnetic-field distributions of the ensemble.

Figure~\ref{fig:pulsarABCD} summarises the post–mass-transfer spin, magnetic-field, and spin-down distributions for all seven model variants. Consistent with Figure~\ref{fig:FGK06}, the ordering of outcomes (fast \texttt{cir-Ed}, slow \texttt{e-Ed}) persists even when $P_0$ is increased by a factor of three (A→B/D). Raising $B_0$ by an order of magnitude (A→C/D) increases the median post–mass-transfer surface field by $\sim$1\,dex and shifts $\dot{P}_\mathrm{spin}$ upward accordingly; yet, even in the highest-$B_0$ runs, \texttt{cir-Ed} pulsars spin down orders of magnitude more slowly than their eccentric counterparts.

Varying the burial scale $\kappa$ from 0.025 to 0.25\,Myr (A→E) modestly increases the residual surface magnetic field, but the \texttt{cir-Ed} vs.\ \texttt{e-Ed} dichotomy remains intact. Similarly, shortening or lengthening the exponential-decay time-scale $\tau$ (A→F/G) modifies the low-$B$ tail of the distribution without shifting \texttt{cir-Ed} systems out of the MSP regime. Across models E–G, variations in $\kappa$ and $\tau$ alter the distribution tails but preserve the channel ordering: the tight cluster of low-$\dot{P}$, sub-10\,ms (i.e. 10$^{-2}$\,s) pulsars arises from continuous, circular RLOF.

\begin{figure*}[!htbp]
  \centering
  \begin{tabular}{@{}cc@{}} 
    \begin{minipage}[t]{0.5\textwidth}
      \includegraphics[width=\linewidth]{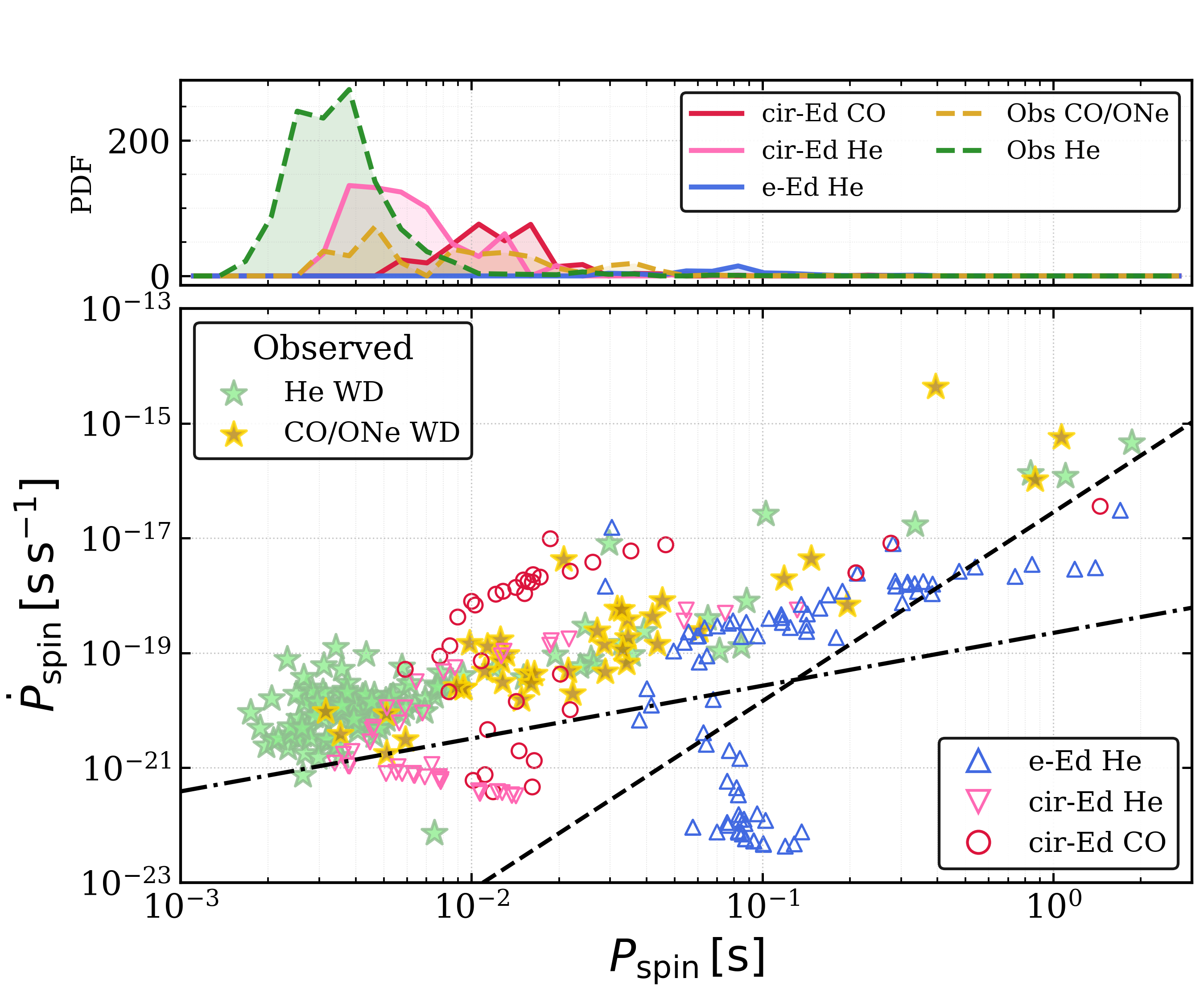}
    \end{minipage} &
    \begin{minipage}[t]{0.5\textwidth}
      \includegraphics[width=\linewidth]{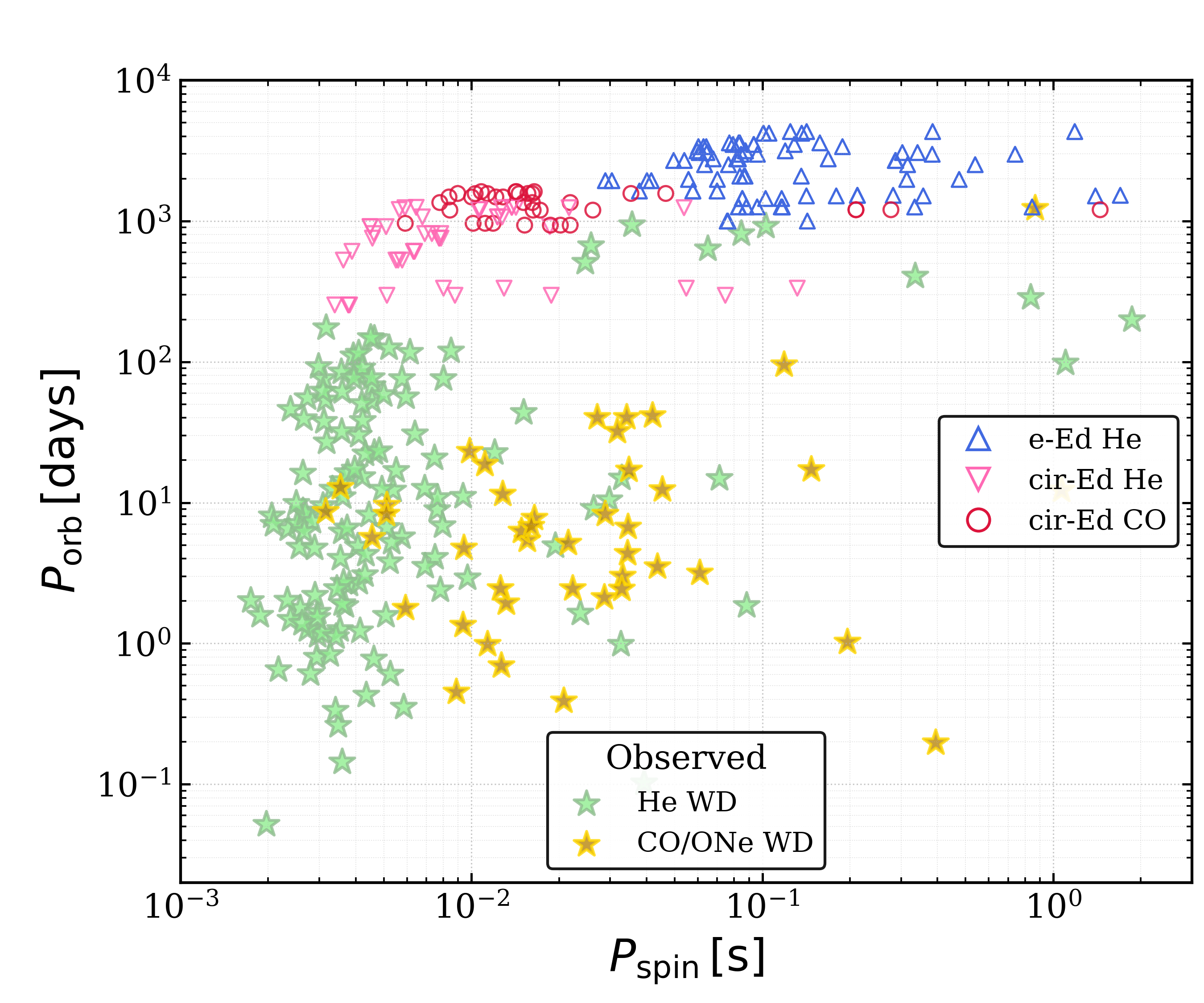 }
    \end{minipage} \\
  \end{tabular}
\caption{Comparison of theory (evolved Gaia NS binaries) with observations of Galactic MSP–WD systems. 
\textbf{Left:} Spin period $P_\mathrm{spin}$ versus spin-down rate $\dot{P}_\mathrm{spin}$ at the end of each simulation for four representative pulsar models (A, C, E, G), chosen to span evolutionary outcomes. The model ensemble contains nearly as many points (21 Gaia binaries $\times$ 4 pulsar models $\times$ 2 mass-transfer prescriptions $\approx 168$) as the observed sample (178 pulsar–WD systems). The \emph{bottom} subpanel shows the $P_\mathrm{spin}$–$\dot{P}_\mathrm{spin}$ distributions: \texttt{e-Ed} (blue, upright triangles) and \texttt{cir-Ed} (red, inverted triangles), over-plotted with observed He-WD (green stars) and CO/ONe-WD (gold circles) systems. The \emph{top} subpanel shows probability density functions (PDFs) for simulated and observed populations, split by WD type and evolutionary channel. While \texttt{e-Ed} broadly reproduces mildly recycled pulsar–He-WD binaries, and \texttt{cir-Ed} aligns better with pulsar–CO-WD systems (and a subset of He-WDs), the bulk of highly recycled He-WD MSPs with $P_\mathrm{spin}\!\lesssim\!5$\,ms are absent from all Gaia-based models. 
\textbf{Right:} $P_\mathrm{spin}$ versus orbital period $P_\mathrm{orb}$ for the same models. Although some wide systems with $P_\mathrm{orb}\!\sim\!10^{3}$\,days appear in both \texttt{e-Ed} and \texttt{cir-Ed}, the evolved Gaia binaries do not produce the short-period pulsar–WD binaries with $P_\mathrm{orb}\!\le\!10^{2}$\,days that are prominent in the observed sample, indicating different initial conditions and/or unstable mass transfer are likely required. The $P_\mathrm{spin}$ distributions for the super-Eddington channels \texttt{e-SpEdcn} and \texttt{e-SpEd100} are effectively similar to \texttt{cir-Ed}, and their $P_\mathrm{orb}$ distributions resemble \texttt{e-Ed}; for clarity, they are not shown.}
\label{fig:PPdotObs2}
\end{figure*}

\subsection{Observational context}\label{sec:obspulsar}

To place our results in a broader pulsar-astronomy framework, we compare our \texttt{MESA} model outcomes in terms of observations of spin period $P_\mathrm{spin}$, spin down rate $\dot{P}_\mathrm{spin}$, orbital period P$_{\rm orb}$ and eccentricity $e$ in Galactic pulsar-WDs.

We illustrate representative outcomes (models A, C, E, and G) in the $P_\mathrm{spin}$–$\dot{P}_\mathrm{spin}$ plane for the \texttt{cir-Ed} channel, using different symbols for He- and CO-WD companions, alongside Galactic field pulsar–WD systems from the ATNF catalogue (left panel of Fig.~\ref{fig:PPdotObs2}). Each model point corresponds to the pulsar at the moment of WD formation. Most He-WD companion models cluster at $P_\mathrm{spin}\lesssim10$\,ms and $\dot{P}_\mathrm{spin}\lesssim10^{-19}$\,s\,s$^{-1}$, whereas CO-WD systems occupy slightly longer periods ($P_\mathrm{spin}\gtrsim20$\,\rm ms) and higher spin-down rates ($\dot{P}_\mathrm{spin}\gtrsim10^{-18}$\,\rm s\,s$^{-1}$). Although all \texttt{cir-Ed} sequences initiate RLOF during Case\,B evolution, lower-mass He-WD progenitors fill their RLs earlier on the first giant branch and sustain stable mass transfer for $\sim10^6$–$10^7$\,yr, delivering more angular momentum and fully recycling the NS. More massive CO-WD progenitors transfer mass for much shorter durations (median $\sim2.8\times10^6$\,yr versus $\sim5.9\times10^6$\,yr for He-WD progenitors), accrete less, and reach only partial recycling. This behaviour mirrors the Galactic MSP–WD population \citep{Tauris2012}.

In the \texttt{e-SpEd100} and \texttt{e-SpEdcn} models, high mass-accretion rates ensure MSP formation, but the net spin-up torque is uncertain. Radiation-pressure–driven winds and disk thickening near the magnetosphere can cap the effective accretion rate, reducing spin-up efficiency and rendering torque prescriptions model-dependent. While NS ULXs show strong spin-up under super-Eddington accretion, these are typically HMXBs; by contrast, LMXBs—well established as MSP progenitors—accrete near Eddington, and direct evidence for MSP formation under sustained super-Eddington accretion remains limited \citep{SpEdAndersson2005,SpEdChashkina2019,SpEdVasilopoulos2020}. Nonetheless, several studies report accretion rates exceeding Eddington by factors of $10^2$–$10^3$ \citep[e.g.][]{Bachetti2014,Israel2017Sci,Chen2017,Tong2024}, and in some cases up to $\sim1200\times$ depending on magnetic-field strength \citep{Inoue2023}, demonstrating that such extremes are physically plausible. Whether these conditions can be sustained in L/IMXB progenitors of MSPs remains unclear. Even in MSP-producing \texttt{e-SpEdcn} models, the resulting NS–WD binaries remain significantly eccentric.

Comparing the spin and orbital periods of the evolved Gaia binaries with those of observed pulsar–WD systems (right panel of Fig.~\ref{fig:PPdotObs2}), the $P_\mathrm{spin}$ distribution partially overlaps both He- and CO/ONe-WD populations. The $P_\mathrm{orb}$ distribution, however, exposes a key shortcoming: the models do not yield the short-period binaries ($\lesssim100$\,days, extending down to hours) that are prominent in the observed sample. Notably, the shortest-$P_\mathrm{orb}$ MSP–WD systems also spin rapidly (a few milliseconds). This discrepancy and correlation point to an alternative formation channel—such as CE evolution with subsequent recycling—for a substantial fraction of Galactic pulsar–WD systems. Gaia is only sensitive to NS-MS binaries in relatively wide orbits \citep{El-Badry2024gen}, which produce wide NS-WD binaries. 

Because Galactic-field MSP–WD binaries are overwhelmingly circular, the tension with eccentric mass-transfer prescriptions may require attention. The mismatch could reflect deficiencies in current eccentric-transfer treatments, but it may also be amplified by strong radio-selection biases against wide, eccentric binaries: rapid periastron passages induce Doppler modulation and ``jerk'', smearing pulsar signals across Fourier bins and suppressing flux by factors $\gamma^2\ll1$ unless acceleration–jerk searches are employed \citep{Bagchi2013,Chattopadhyay2021}. Partially recycled pulsars formed via impulsive \texttt{e-Ed} episodes also spin down more rapidly toward the radio death line, further reducing detectability. On the theoretical side, population synthesis with \texttt{POSYDON} predicts that $\approx70\%$ of NS–WD progenitors undergo CE evolution rather than stable RLOF \citep{Chattopadhyay2025inprep}. Under standard initial orbital distributions, Gaia-like systems arise only in a narrow region of parameter space. Thus, although enforced circular RLOF naturally reproduces the low eccentricities of the MSP–WD population, eccentric channels may be underrepresented due to low formation rates, rapid spin-down, radio-selection effects, or alternative pathways (e.g., cluster ejections). Caution is warranted when interpreting circularisation in \texttt{cir-Ed}, since the resulting $P_\mathrm{orb}$ remains larger than would be expected from a CE channel.

Taken together, these results suggest that the Gaia NS–MS binaries are unlikely to be the dominant progenitors of the Galactic MSP–WD population. Multiple evolutionary pathways probably contribute, with different channels populating distinct regions of $P_\mathrm{spin}$–$P_\mathrm{orb}$ space. Considering $P_\mathrm{spin}$, $P_\mathrm{orb}$, and $e$ jointly, the absence of very short orbital periods in our models—despite their prevalence in the observations—argues for markedly different initial mass–orbit configurations and, in many cases, unstable mass transfer being essential to producing a significant fraction of Galactic MSPs.

Observationally, Galactic WD–NS systems show that the fastest pulsars (small $P_{\rm spin}$) tend, on average, to lie in longer orbits. Using ATNF catalogued values, helium–WD systems host faster-spinning pulsars (median $P_\mathrm{spin}$=4.1\,ms) and longer orbital periods (median $P_{\rm orb}$=9.6\,days) than carbon-oxygen/oxygen-neon –WD systems (median $P_\mathrm{orb}$=21.5\,ms, $\sim\!5\times$ slower; $P_{\rm orb}$=6.3\,days, $\sim\!1.5\times$ shorter) \citep{Manchester2005}. This makes the contrast between the \texttt{e-Ed} and fully conservative \texttt{e-SpEdcn} cases particularly informative: strongly recycled pulsars with helium–WD companions are observed in longer orbits, whereas non-conservative stable mass transfer naively predicts shorter (or only weakly widened) orbits. Reconciling these trends keeping to the exact same eccentric mass transfer prescription as \citet{Sepinsky2009} likely requires additional evolutionary phases (e.g., common-envelope episodes) and recognition that Gaia’s NS–MS progenitors may not follow the same forward channels as the systems that populate the Galactic WD–NS sample \citep{Chattopadhyay2025inprep}.

\section{Summary} \label{sec:summary}
We model the future evolution of 21 Gaia NS–MS binaries using 1D stellar evolution with \texttt{MESA}, incorporating both traditional artificially circularized mass-transfer and eccentric prescriptions. We then connect these outcomes to the observed Galactic pulsar–white dwarf population through pulsar spin and magnetic-field evolution modelling. Our main findings are as follows

\begin{enumerate}
    \item \textbf{Orbital period and eccentricity:} The two mass‐transfer prescriptions produce systematically different final orbits. In eccentric models, systems expand to final $P_\mathrm{orb}$ values $\sim3$–$4\times$ (median) their initial values, while $e$ typically grows to $\sim1.5\times$ the initial value. In the circular channel, tides are assumed to circularize the orbit at RLOF onset: $P_\mathrm{orb}$ often shrinks initially and then widens during stable transfer, yielding final periods typically $\sim1.5\times$ the initial value (only four cases end slightly shorter). The resulting circularized NS–WD binaries span $200\,\mathrm{days}\!\lesssim\!P_\mathrm{orb}\!\lesssim\!1000\,\mathrm{days}$ and overlap a subset of the Galactic pulsar–WD population. Across both channels, the orbital angular‐momentum budget is dominated by mass loss (and tides) before/during RLOF; after detachment, mass loss dominates, with magnetic braking and gravitational radiation being negligible for these wide systems.
    
    \item \textbf{Mass transfer:} Eccentric models exhibit short‐lived, high peaks $(-5 \lesssim \log_{10}\dot M/\mathrm{M_\odot\,yr^{-1}} \lesssim -3)$ confined to periastron and lasting $\lesssim10^6$\,yr. Circular models sustain $\log_{10}\dot M/\mathrm{M_\odot\,yr^{-1}} \!\sim\!-6$ for up to $10^7$\,yr. Consequently, Eddington‐limited circular runs typically accrete $\sim0.1\,\mathrm{M_\odot}$ onto the NS—about an order of magnitude above the $\sim(2$–$5)\!\times\!10^{-2}\,\mathrm{M_\odot}$ in eccentric runs. As variations, we also explored super‐Eddington cases, allowing up to $100\times$ Eddington and, in fully conservative limits (effective rates approaching $\sim3000\times$).
    
    \item \textbf{WD outcomes and HR diagram behaviour:} The longer, steadier stripping in the circular channel lets some donors exhaust most of their hydrogen envelopes and grow cores past the helium ignition threshold ($\simeq0.45\,\mathrm{M_\odot}$), yielding CO WDs and, in some cases, shell flash loops in the HR diagram. This occurs in nine out of 21 circular runs. All eccentric runs detach too early for core He ignition and produce He WDs. Thus, the circular channel yields a mixed He/CO WD population, whereas the eccentric channel yields exclusively He WDs.
    
    \item \textbf{Recycling and millisecond pulsars:} Exploring uncertainties in pulsar birth spin/field and in decay parameters ($\tau$ and $\kappa$), we find that under Eddington‐limited accretion, circular RLOF efficiently produces fully recycled MSPs ($P_\mathrm{spin}\!\sim$few–$30$\,ms). Eccentric transfer typically yields only partially or mildly recycled pulsars ($P_\mathrm{spin}\!\gtrsim\!50$\,ms, up to $\sim1$\,s). Allowing super‐Eddington accretion ($\lesssim100\times$) can produce MSPs even in eccentric runs, but torque efficiencies there are model‐dependent.
    
    \item \textbf{Comparison with observed pulsar-WDs:} The modelled $P_\mathrm{spin}$ distribution shows partial overlap with both He‐ and CO/ONe–WD systems, but the $P_\mathrm{orb}$ distribution reveals a key tension: our forward evolution from Gaia NS–MS binaries does not produce the numerous short‐period ($\lesssim100$\,day, down to hours) pulsar–WD binaries seen in the Galaxy. This strongly points to either vastly different initial mass and orbital parameters and/or additional formation channels—e.g., common‐envelope evolution—being required to populate the short‐$P_\mathrm{orb}$ tail. Gaia is not sensitive to short-period NS-MS binaries, whose orbits are too small on the plane of the sky to astrometrically detect.
\end{enumerate}

Overall, the forward evolution of Gaia NS–MS binaries with artificially circularized RLOF yields efficient spin-up and near-circular orbits, but fails to reproduce the bulk of short–orbital-period MSP–WD systems. Periastron‐limited eccentric transfer leaves systems wide and eccentric and, under Eddington‐limited accretion, typically only mildly recycled; MSPs in eccentric orbits are possible only if the NS accretes at $\gtrsim100\times$ the Eddington limit or over. The weak correspondence to the bulk MSP–WD population may indicate that the Gaia sample traces a rarer, largely CE‐free channel, or that radio selection disfavours wide/eccentric pulsars.


\section{acknowledgments}
We are grateful to Jeff Andrews, Vivek Venkatraman Krishnan and Manjari Bagchi for their useful comments. D.C. thanks the Gordon and Betty Moore Foundation for funding this research through Grant GBMF12341, and S.G. through GBMF12341 and GBMF8477.
K.A.R thanks the LSSTC Data Science Fellowship Program, which is funded by LSST Corporation, NSF Cybertraining Grant No.1829740, the Brinson Foundation, and the Gordon and Betty Moore Foundation. V.K. is supported by the Gordon and Betty Moore Foundation (grant awards GBMF8477 and GBMF12341), through a Guggenheim Fellowship,
and the D.I. Linzer Distinguished University Professorship fund. K.EB. is supported by NSF grant
AST-2307232. A.T.
is supported by NSF grants AST-2009884, AST-2107839,
AST-1815304, AST-1911080, AST-2206471, AST-2407475,
and OAC-2031997, and by NASA grants 80NSSC22K0031,
80NSSC22K0799, 80NSSC18K0565, and 80NSSC21K1746.

The computations were performed at Northwestern University on the Trident computer cluster (funded by the GBMF8477 award). 
This research was supported in part through the computational resources and staff contributions provided for the Quest high performance computing facility at Northwestern University which is jointly supported by the Office of the Provost, the Office for Research, and Northwestern University Information Technology.

\section{Gold Open Access}

The data produced for this work will be freely available upon reasonable request to the corresponding author.

\bibliography{sample631}{}
\bibliographystyle{aasjournal}

\end{document}